\documentclass[10pt,twoside]{article}

\usepackage{a4}
\usepackage{amsmath}
\usepackage{amssymb}
\usepackage{amsthm}
\usepackage{amsfonts}
\usepackage{amstext}
\usepackage{amsopn}
\usepackage{amsxtra}
\usepackage{graphicx}
\usepackage[latin1]{inputenc}
\usepackage{eufrak}
\usepackage{dsfont}
\usepackage{mathrsfs}
\newcommand\R{{\ensuremath {\mathbb R} }}
\newcommand\C{{\ensuremath {\mathbb C} }}

\newcommand\Z{{\ensuremath {\mathbb Z} }}

\newcommand\1{{\ensuremath {\mathds 1} }}
\renewcommand\S{\mathcal{S}}
\renewcommand\phi{\varphi}

\newcommand{\alp}{\boldsymbol{\alpha}}
\newcommand{\bPsi}{\boldsymbol{\Psi}}
\newcommand{\gH}{\mathfrak{H}}
\newcommand{\gS}{\mathfrak{S}}
\newcommand{\CJ}{\mathscr{C}}
\newcommand{\wto}{\rightharpoonup}
\newcommand{\cP}{\mathcal{P}}
\newcommand{\cD}{\mathcal{D}}
\newcommand{\sgn}{{\rm sgn}}

\newcommand\ii{{\ensuremath {\infty}}}
\newcommand\pscal[1]{{\ensuremath{\langle #1 \rangle}}}
\newcommand{\norm}[1]{ \left| \! \left| #1 \right| \! \right| }

\def\tr{\mathop{\rm tr}\nolimits} 
\newcommand{\E}{\mathcal{E}}
\newcommand{\F}{\mathcal{F}}
\newcommand{\T}{\mathcal{T}}
\newcommand{\A}{\mathcal{A}}
\newcommand{\B}{\mathcal{B}}
\newcommand{\cC}{\mathcal{C}}
\newcommand{\cQ}{\mathcal{Q}}
\newcommand{\bT}{\mathbb{T}}
\newcommand{\cG}{\mathcal{G}}

\newtheorem{theorem}{Theorem}
\newtheorem{lemma}{Lemma}

\newtheorem{remark}{Remark}

\title{\bf The Mean-Field Approximation in Quantum Electrodynamics.\\ The no-photon case.}
\author{\bf Christian HAINZL, Mathieu LEWIN \& Jan Philip SOLOVEJ}
\date{August 26, 2005}
\begin{document}
\maketitle

\begin{center}
University of Copenhagen, Department of Mathematics,

Universitetsparken 5, 2100 Copenhagen \O, DENMARK.

\bigskip

E-mail: {\it hainzl, lewin, solovej@math.ku.dk}
\end{center}

\begin{abstract}
We study the mean-field approximation of Quantum Electrodynamics, by means of a thermodynamic limit. The QED Hamiltonian is written in Coulomb gauge and does not contain any normal-ordering or choice of bare electron/positron subspaces. Neglecting photons, we define properly this Hamiltonian in a finite box $[-L/2;L/2)^3$, with periodic boundary conditions and an ultraviolet cut-off $\Lambda$. We then study the limit of the ground state (i.e. the vacuum) energy and of the minimizers as $L$ goes to infinity, in the Hartree-Fock approximation.

In case with no external field, we prove that the energy per volume converges and obtain in the limit a translation-invariant projector describing the free Hartree-Fock vacuum. We also define the energy per unit volume of translation-invariant states and prove that the free vacuum is the unique minimizer of this energy.

In the presence of an external field, we prove that the difference between the minimum energy and the energy of the free vacuum converges as $L$ goes to infinity. We obtain in the limit the so-called Bogoliubov-Dirac-Fock functional. The Hartree-Fock (polarized) vacuum is a Hilbert-Schmidt perturbation of the free vacuum and it minimizes the Bogoliubov-Dirac-Fock energy.
\end{abstract}

\maketitle





\section{Introduction}
In Coulomb gauge and when photons are neglected, the Hamiltonian of Quantum Electrodynamics (QED) reads formally \cite{Hei,HE,Se,Sch1,BD}
\begin{equation}
\label{Hamiltonian01}
\mathbb{H}^\phi = \int \Psi^*(x) D^0 \Psi(x) \,dx -\int\phi(x){\rho}(x)dx+ \frac{\alpha}2 \iint \frac{\rho(x)\rho(y)}{|x-y|}dx\,dy
\end{equation}
where $\Psi(x)$ is the second-quantized field operator satisfying the usual anti-commutation relations, and $\rho(x)$ is the density \emph{operator}
\begin{equation}
\label{Rho1}
\rho(x)=\frac{\tr[\Psi^*(x),\Psi(x)]}{2}=\frac{\sum_{\sigma=1}^4\left\{\Psi^*(x)_\sigma\Psi(x)_\sigma-\Psi(x)_\sigma\Psi^*(x)_\sigma\right\}}{2},
\end{equation}
where $\sigma$ is the spin variable.
In \eqref{Hamiltonian01}, $D^0=-i\alp\cdot\nabla+m_0\beta$ is the usual free Dirac operator \cite{Thaller} ($m_0>0$ is the \emph{bare} mass of the electron),
$\alpha$ is the \emph{bare} Sommerfeld fine structure constant and $\phi$ is the external potential. We have chosen a system of units such that $\hbar=c=1$.

In QED, one main issue is the minimization of the Hamiltonian \eqref{Hamiltonian01}. For the study of systems like atoms or molecules, one usually uses the Born-Oppenheimer approximation and treats the nuclei as fixed and time-independent external sources, justifying the introduction of the external field $\phi$. A global minimizer of $\mathbb{H }^\phi$ would then be interpreted as the free vacuum if $\phi=0$, or the polarized vacuum when $\phi\neq0$. On the other hand, a minimizer in the $N$th charge sector (in a sense to be made more precise) would describe a bound state with particles.

However, none of these minimization problems make sense \emph{a priori}, since $\mathbb{H }^\phi$ cannot be bounded from below. This old problem \cite{D1,D2} is due to the fact that the free Dirac operator $D^0$ has a negative essential spectrum: $\sigma(D^0)=(-\ii;-m_0]\cup[m_0;\ii)$. Our goal in this paper will be to use a thermodynamic limit in order to give a precise mathematical meaning to the minimization of $\mathbb{H }^\phi$ in the mean-field approximation, that is to say when it is restricted to Hartree-Fock (HF) states. We shall in particular be able to define properly the Hartree-Fock global minimizer (i.e. the vacuum), with or without an external field $\phi$.

We have neglected photons in \eqref{Hamiltonian01}. This approximation is physically justified when studying the free vacuum (i.e. when $\phi=0$). In full QED, the expectation value of the photon field $\langle\vec{A}(x)\rangle$ vanishes for the (non Hartree-Fock) free vacuum, see e.g. \cite{Sch2,Sch3,PV}. But photons should be included for a complete treatment of the external field case.

We emphasize that \eqref{Hamiltonian01} does not contain any normal-ordering or notion of (bare) electrons and positrons: $\Psi(x)$ can annihilate electrons of negative kinetic energy. Indeed, the distinction between electrons and positrons should be a result of the theory and not an input. The commutator used in the formula \eqref{Rho1} of $\rho(x)$ is a kind of renormalization, independent of any reference. It is due to Heisenberg \cite{Hei} (see also \cite[Eq. $(96)$]{Pauli} and \cite[Eq. $(38)$]{Dy1}), and has been widely used by Schwinger (see \cite[Eq. $(1.14)$]{Sch1}, \cite[Eq. $(1.69)$]{Sch2} and \cite[Eq. $(2.3)$]{Sch3}) as a necessity for a covariant formulation of QED. More precisely, the Hamiltonian $\mathbb{H}^\phi$ possesses the interesting property of being invariant under charge conjugation since the following relations  hold formally
$$\CJ\rho(x)\CJ^{-1}=-\rho(x),\qquad \CJ\mathbb{H}^\phi \CJ^{-1}=\mathbb{H}^{-\phi},$$
where $\CJ$ is the charge conjugation operator acting on the Fock space (details will be given later on).
Notice that the use of a commutator in the same way as \eqref{Rho1} for the kinetic energy term would have no effect since the Dirac matrices are trace-less:
$$\frac12\int \tr[\Psi^*(x), D^0 \Psi(x)] \,dx=\int \Psi^*(x) D^0 \Psi(x) \,dx.$$

In the mean-field approximation, one restricts the QED Hamiltonian \eqref{Hamiltonian01} to Hartree-Fock states in the underlying Fock space. These states $|\Omega\rangle$ have the property of being totally described, up to a phase, by the two-point  function $P(x,y) = \langle\Omega|\Psi^*(x)\Psi(y)|\Omega\rangle$, also called the \emph{density matrix of $|\Omega\rangle$}. In the usual Hartree-Fock theory, the energy only depends on the orthogonal projector $P$. This is also true in QED but, due to the specific definition of the density operator $\rho(x)$ in \eqref{Rho1}, the energy will be more easily expressed in terms of the \emph{renormalized density matrix}
\begin{equation}
\gamma:=\frac{P-P^\perp}{2}=P-1/2
\label{def_gamma_intro}
\end{equation}
where we have introduced $P^\perp=1-P$, or equivalently
$$\gamma(x,y):=\langle\Omega|\left(\frac{[\Psi^*(x),\Psi(y)]}{2}\right)|\Omega\rangle.$$
The operator $\gamma$ and the associated projector $P=\gamma+1/2$ will be our main objects of concern in this paper. Notice that $\gamma$ is closely related to the Feynman propagator $S_F$ taken at equal times \cite{BD}, a widely used object in QED (for the true non-HF QED vacuum):
$$\gamma(x,y)=-iS_F(x,y;t_x=t_y) \beta.$$

In order to give a precise mathematical meaning to the minimization of $\mathbb{H }^\phi$ in the Hartree-Fock class, we proceed as follows: we first define properly $\mathbb{H }^\phi$ in a finite box in space $\cC_L=[-L/2;L/2)^3$ and with an ultraviolet cut-off $\Lambda$ in the Fourier domain. For the sake of simplicity, we also use periodic boundary conditions on $\cC_L$, i.e. we work on the torus $\bT_L:=\R^3/(L\Z^3)$. Since the problem becomes finite dimensional, the minimization of this well-defined Hamiltonian restricted to Hartree-Fock states makes sense. Let us denote by $E_L(\phi)$ the minimum energy in the presence of the external field $\phi$. The main goal of this paper will be to study the \emph{thermodynamic limit} $L\to\ii$, i.e. the behavior of both the minimizers and the minimum energy $E_L(\phi)$ when the size of the box grows.

Our states will always be represented by their density matrix $P$ or their renormalized density matrix $\gamma=P-1/2$, defined in \eqref{def_gamma_intro}. We shall prove that, as $L$ goes to infinity, the sequence of global minimizers indeed converges to a state defined on the whole space, which will be interpreted as the Hartree-Fock global minimizer of the QED Hamiltonian.

Moreover, our method will also allow us to define the energy of these states. We have to study separately the free case $\phi=0$ and the external field case $\phi\neq0$. In the free case, we shall define properly the \emph{energy per unit volume} of translation-invariant states; the HF free vacuum will be the unique minimizer of this energy. In the external field case, we will obtain at the limit the Bogoliubov-Dirac-Fock (BDF) energy \cite{CI} which is the energy of the state, measured with respect to the energy of the free vacuum. The HF polarized vacuum will be a minimizer of this energy. Therefore, the BDF theory is the appropriate model for the study of Hartree-Fock states in QED in the presence of an external potential.

These limits are studied for a fixed ultraviolet cut-off $\Lambda$. Our objects will be logarithmically divergent in $\Lambda$. The elimination of this last divergence would require a \emph{renormalization}, which we do not address in details here.

\medskip

Let us now describe the results of this paper in more details.

We start with the free case $\phi=0$. First, we prove the existence, for each $L$, of a \emph{unique global minimizer} of $E_L(0)$, $\gamma_L^0=\cP^0_L-1/2$, which is moreover invariant under translations.
We then study the energy per unit volume and prove that
\begin{equation}
\lim_{L\to\ii}\frac{E_L(0)}{L^3}=\min_{\substack{\gamma=f(-i\nabla),\\ -1/2\leq\gamma\leq 1/2}}\T(f).
\label{energ_vol0}
\end{equation}
The energy $\T$ is a natural functional defined for translation-invariant operators only, $\gamma=f(-i\nabla)$, by
\begin{equation}
\T(f)=\frac{1}{(2\pi)^{3}}\int_{B(0,\Lambda)}\tr_{\C^4}[D^0(p)f(p)]dp-\frac{\alpha}{(2\pi)^5}\iint_{B(0,\Lambda)^2}\frac{\tr_{\C^4}[f(p)f(q)]}{|p-q|^2}dp\,dq,
\label{def_fn_F0}
\end{equation}
where we recall that $\Lambda$ is the ultraviolet cut-off. The functional $\T$ indeed can be interpreted as the \emph{energy per unit volume} of translation-invariant states.

We prove that, when $L\to\ii$, the sequence $(\gamma^0_L)$ converges in some sense to the \emph{unique translation-invariant minimizer}  of the r.h.s. of \eqref{energ_vol0}, denoted by $\gamma^0=\cP^0_--1/2$ where $\cP^0_-$ is an orthogonal projector. This state is interpreted as the HF free vacuum. Writing the Euler-Lagrange equation satisfied by $\gamma^0$, we obtain the self-consistent equation
\begin{equation}
\label{equation_scf0}
\left\{\begin{array}{l}
\displaystyle \gamma^0=-\frac{\sgn(\cD^0)}{2},\\
\displaystyle \cD^0=D^0-\alpha\frac{\gamma^0(x,y)}{|x-y|}.
\end{array} \right.
\end{equation}
Written in terms of $\cP^0_-=\gamma^0+1/2$, this equation reads
\begin{equation}
\cP^0_-=\chi_{(-\ii;0)}(\cD^0),
\label{equation_scf0b}
\end{equation}
which corresponds to the usual Dirac picture that the density matrix of the free vacuum should be the projector associated with the negative part of the spectrum of a translation-invariant Dirac operator.

Let us emphasize that in many works, the free vacuum is \emph{assumed} to be represented by the free Dirac projector
$$P^0_-:=\chi_{(-\ii;0)}(D^0),$$
following thereby ideas of Dirac \cite{D1,D2} (see, e.g. \cite{CI,CIL,KS,KS2,BBHS,HLS1,HLS2}).
The true Hartree-Fock vacuum $\cP^0_-$ obtained in this paper is different from $P^0_-$, except when $\alpha=0$. The interpretation is that, contrary to $P^0_-$, the self-consistent interactions between the virtual particles of the vacuum are taken into account via the term $\alpha\gamma^0(x-y)|x-y|^{-1}$ in \eqref{equation_scf0}.

The self-consistent operator $\cD^0$ defined in \eqref{equation_scf0} has an interesting special form. Using the usual notation $p=-i\nabla$, it can be written
$$\cD^0(p)=\alp\cdot p\frac{g_1(|p|)}{|p|}+\beta g_0(|p|)$$
which means that, due to \eqref{equation_scf0},
\begin{equation}
\gamma^0(p)=-\frac{g_1(|p|)}{2|p|\sqrt{g_1(|p|)^2+g_0(|p|)^2}}\alp\cdot p-\frac{g_0(|p|)}{2\sqrt{g_1(|p|)^2+g_0(|p|)^2}}\beta.
\label{scf_gamma}
\end{equation}
In QED, the Feynman propagator $S_F$ is often expressed using the Källén-Lehmann representation \cite{Ka,Leh,BD}, based on relativistic invariances. Although our model is not fully relativistically invariant (we discard photons and use an ultraviolet cut-off $\Lambda$) and is only defined in the mean-field approximation, our solution \eqref{scf_gamma} has exactly the form which may be derived from the Källén-Lehmann representation for the equal time propagator. In four-dimensional full QED, a self-consistent equation similar to \eqref{equation_scf0} is well-known and used. These so-called Schwinger-Dyson equations \cite{Sch4,Dy2} have been approximately solved for the free vacuum case first by Landau {\it et al.} in \cite{Lan81,Lan89}, and then by many authors (see, e.g., \cite{mass1,mass2,mass3}).

In \cite{LSie}, Lieb and Siedentop arrived at the same equation \eqref{equation_scf0} with totally different arguments from ours. In particular, they did not derive the equation \eqref{equation_scf0} as the Euler-Lagrange equation associated with a minimization problem, but rather looked for a self-consistent normal-ordering in a free Hamiltonian. They proved the existence of a solution of \eqref{equation_scf0} by means of a fixed-point method, valid under a restrictive condition of the form $\alpha\log\Lambda<C$ where $\Lambda$ is the ultraviolet cut-off. Our proof is completely different since it proceeds by minimizing the functional $\T$. This enables us to prove the existence of a solution to \eqref{equation_scf0} without any constraint linking $\alpha$ and $\Lambda$.

Let us remark that since $\gamma^0$ is translation-invariant, the associated \emph{density of charge}, formally defined by $\rho_{\gamma^0}(x)=\tr_{\C^4}\gamma^0(x,x)$, is indeed a constant. A consequence of the special form \eqref{scf_gamma} of $\gamma^0$ is that $\rho_{\gamma^0}(x)\equiv0$, the Dirac matrices being trace-less. Therefore, in this formalism and thanks to \eqref{Rho1}, the free vacuum has no local density of charge, which is physically comforting.

We know that $\gamma^0$ is a minimizer of the energy $\T$ among other translation-invariant operators. Since however $\gamma^0_L$ is for any $L$ a \emph{global} minimizer, we will also prove (in a sense to be made more precise later) that the limiting $\gamma^0$ is also a minimizer of $\mathbb{H}^0$ under local Hilbert-Schmidt perturbations. Our study therefore shows that although one cannot give a meaning to the energy of the free HF vacuum, one can give a sense of being a minimizer of the Hartree-Fock energy, either among translation-invariant operators, or under Hilbert-Schmidt perturbations.

\medskip

We then study the external field case $\phi\neq0$. We prove that the energy measured with respect to the free energy has a limit:
\begin{equation}
\lim_{L\to\ii}\left\{E_L(\phi)-E_L(0)\right\}=\min_{\substack{\gamma-\gamma^0\in\gS_2,\\-1/2\leq \gamma\leq 1/2}}\E_{\rm BDF}^\phi(\gamma-\gamma^0).
\label{limit_ext_intro}
\end{equation}
Here $\E_{\rm BDF}^\phi$ is the so-called Bogoliubov-Dirac-Fock (BDF) energy \cite{CI,Chaix,BBHS,HLS1,HLS2}, associated with the free vacuum $\gamma^0$. It measures the energy of a HF state $|\Omega\rangle$ with renormalized density matrix $\gamma$, relatively to the energy of the free vacuum $|\Omega_0\rangle$, whose renormalized density matrix $\gamma^0$ has been defined previously, i.e.
\begin{equation}
\E_{\rm BDF}^\phi(\gamma-\gamma^0) = \mbox{``}\langle\Omega|\mathbb{H}^\phi|\Omega\rangle-\langle\Omega_0|\mathbb{H}^0|\Omega_0\rangle\mbox{''}.
\end{equation}
The BDF energy reads formally
\begin{multline}
\E^\phi_{\rm BDF}(Q)=\tr[\cD^0(\cP^0_+Q\cP^0_++\cP^0_-Q\cP^0_-)]-\int\phi(x)\rho_{Q}(x)dx\\+\frac\alpha2\iint\frac{\rho_Q(x)\rho_Q(y)}{|x-y|}dx\,dy-\frac\alpha2\iint\frac{|Q(x,y)|^2}{|x-y|}dx\,dy,
\label{BDF_intro2}
\end{multline}
where $\rho_Q(x)=\tr_{\C^4}(Q(x,x))$ is the charge density, a well-defined object when $Q\in\gS_2$, thanks to the ultraviolet cut-off $\Lambda$ \cite{HLS1}.

For any $L$, we prove the existence of a Hartree-Fock global minimizer for $E_L(\phi)$, $\gamma_L=\cP_L-1/2$. Then we show that $\gamma_L-\gamma_L^0=\cP_L-\cP_L^0$ converges, in some sense, to a global minimizer $\bar Q=\bar\gamma-\gamma^0$ of the BDF energy \eqref{BDF_intro2}.
Hence, the solution $\bar\gamma=\gamma^0+\bar Q$ of the r.h.s. of \eqref{limit_ext_intro} is a Hilbert-Schmidt perturbation of the free $\gamma^0$ and solves the self-consistent equation
\begin{equation}
\label{Equation_gamma_02}
\displaystyle\bar\gamma=-\frac{\sgn(\bar\cD)}{2}
\end{equation}
with
\begin{eqnarray}
 \bar\cD & = & \displaystyle\cD^0-\phi+\alpha\rho_{(\bar\gamma-\gamma^0)}\ast\frac{1}{|\cdot|}-\alpha\frac{(\bar\gamma-\gamma^0)(x,y)}{|x-y|}\nonumber\\
 & = & \displaystyle D^0-\phi+\alpha\rho_{\bar\gamma}\ast\frac{1}{|\cdot|}-\alpha\frac{\bar\gamma(x,y)}{|x-y|},\label{form_D_intro}
\end{eqnarray}
where we have used the definition of $\cD^0$ in \eqref{equation_scf0} and $\rho_{\gamma^0}\equiv0$. Written in terms of the projector $\bar\cP_-=\bar\gamma+1/2$, \eqref{Equation_gamma_02} can be written
\begin{equation}
\label{EquationP_02}
\displaystyle\bar\cP_-=\chi_{(-\ii;0)}(\bar\cD),
\end{equation}
which one more time corresponds to Dirac's interpretation that the density matrix of the vacuum should be the projector associated with the negative spectrum of an effective Dirac operator.
We emphasize that, due to \eqref{form_D_intro}, the self-consistent equation \eqref{Equation_gamma_02} does not depend on the reference $\gamma^0$ used for the definition of the BDF energy.

Self-consistent equations for relativistic Hartree-Fock states are well-known in QED. We refer the reader for instance to \cite[Eq. $(4)$]{RGA} which is exactly equivalent to \eqref{EquationP_02}, and to \cite{RG,ED,GT,DH,Hamm} for related studies. In \cite{ED,Engel}, similar equations are obtained in the relativistic density functional theory.

Notice that the BDF energy \eqref{BDF_intro2} can also be used in the free case $\phi=0$. Similarly to \cite{CIL,BBHS,HLS1}, we shall prove in this case that its minimizer is $0$, which precisely means that the free vacuum $\gamma^0$ is a minimizer of $\mathbb{H}^0$ under Hilbert-Schmidt perturbations.

The Bogoliubov-Dirac-Fock energy has been introduced first by Chaix-Iracane in \cite{CI,Chaix}. They also study a mean-field approximation of QED, but start with a different Hamiltonian: taking the free Dirac projector $P^0_-$ as a definition of the free vacuum, they used a Hamiltonian normal-ordered with respect to $P^0_-$. In \cite{CIL}, Chaix, Iracane and Lions proved that their free vacuum $P^0_-$ is stable when no external field is present. Our BDF energy \eqref{BDF_intro2} is similar to the one of Chaix-Iracane, but $P^0_-$ and $D^0$ have been replaced by $\cP^0_-$ and $\cD^0$.

The BDF model of Chaix-Iracane has been studied by Bach, Barbaroux, Helffer and Siedentop in \cite{BBHS}, in which a mathematical setting without any ultraviolet cut-off is provided when $\phi=0$, to prove the stability of the free vacuum $P^0_-$ under trace-class perturbations. The case of an external field is studied as well, but the vacuum polarization is neglected, leading to a totally different model.

A rigorous framework for the study of the external field case has been recently provided by Hainzl, Lewin and Séré in \cite{HLS1,HLS2}. There, an ultraviolet cut-off is introduced and the energy is defined on a set of Hilbert-Schmidt operators which are not necessarily trace-class. Starting from the Hamiltonian of Chaix and Iracane, it is proved that the corresponding BDF energy is bounded from below and that it possesses a minimizer. It satisfies an equation similar to \eqref{EquationP_02}, but which still explicitly depends on the chosen reference $P^0_-$ \cite[Eq. $(6)$]{HLS1}. We shall rely heavily on \cite{HLS1,HLS2}. In particular, we shall generalize the results of \cite{HLS2} to the new BDF energy \eqref{BDF_intro2} and obtain the existence of a minimizer satisfying our equation \eqref{EquationP_02}. For a time-dependent study of the BDF model of Chaix-Iracane, we refer to \cite{HLSp}.

\medskip

Summarizing our results, we have been able to give a meaning to the Hartree-Fock approximation in no-photon QED, by means of a thermodynamic limit. The free vacuum has a renormalized density matrix  $\gamma^0=\cP^0_--1/2$, which is a translation-invariant operator, solution of \eqref{equation_scf0}. It is a minimizer among other translation invariant operators of the energy per unit volume $\T$ defined in \eqref{def_fn_F0} and as well a minimizer of $\mathbb{H}^0$ under Hilbert-Schmidt perturbations. In the presence of an external potential, the HF polarized vacuum is a Hilbert-Schmidt perturbation $\bar\gamma=\gamma^0+\bar Q$ of the free $\gamma^0$. The operator $\bar Q$ minimizes the BDF functional \eqref{BDF_intro2} and $\bar\gamma$ solves \eqref{Equation_gamma_02}.

This work shows that the BDF functional \eqref{BDF_intro2} is the appropriate tool for the study of Hartree-Fock states in QED, in the presence of an external potential.
Although the present paper only addresses the case of the vacuum, our method should be applicable to the case of atoms and molecules. In that case, one would have to minimize the BDF energy in a sector of fixed charge $N$.

As mentioned above, the theory is still divergent in the cut-off $\Lambda$. This phenomenon is also encountered in full QED, where the concepts of \emph{mass and charge renormalization} \cite{Dy2,BD} are used to get finite physical quantities.
Mass and charge renormalization will usually not remove the divergences in
the propagator $S_F$ or in the equal time propagator $\gamma$. This is not
surprising since $\gamma$ as defined in \eqref{def_gamma_intro} does not represent a physical
quantity, it is not the propagator of a physical particle. The physical
electron (at rest) is represented by the lowest energy state in the charge
one sector (the positron corresponds to the charge -1 sector).  In
particular, the rest mass of the physical particle is the lowest energy in
the charge one sector. The {\it physical electron} is not just the bare
particle, but includes also a cloud of virtual particles. The relation
between the bare propagator $S_F$ and the propagator for the physical
particle is in the physics literature often expressed by what is called a
\emph{wavefunction renormalization}. We will discuss the concept of
renormalization in Section \ref{sec_ren}.  As we will explain, wavefunction
renormalization is conceptually different from charge and mass
renormalization. A completely different notion of wavefunction
renormalization, which also leads to a different notion of physical mass,
was introduced in \cite{LSie}.

Of course, this study only applies to the Hartree-Fock approximation, the general case being much more difficult in our point of view. In particular, we do not know if the use of a normal-ordered Hamiltonian with respect to the free electron/positron spaces defined by ${\cP^0_-}$ as proposed in \cite{LSie}, is physically relevant. By definition, the normal-ordering procedure takes a projector (i.e. a  Hartree-Fock state) as a reference, whereas the true vacuum is known to be a non Hartree-Fock state. The study of a thermodynamic limit for the full QED model could be a better approach in the quest of a non-perturbative formulation of QED.

\medskip

The paper is organized as follows. In the next section, we define properly the models and state our main theorems. We start by giving a meaning to the QED Hamiltonian in the box $\cC_L$ and formulate the thermodynamic problem mathematically. Then, we define the models on the whole space that will be obtained as thermodynamic limits: the energy $\T$ of translation-invariant projectors and its minimizer $\gamma^0=\cP^0_--1/2$, and the associated Bogoliubov-Dirac-Fock model. Finally, we state our main results concerning the thermodynamic limit. For the sake of clarity, we have brought all the proofs together at the end of the paper, in Section \ref{proofs}. Section \ref{sec_ren} is devoted to the discussion of the renormalization problem.

\bigskip
\noindent {\bf Acknowledgment.} {\it The authors are thankful to Jan
Derezinski for useful advice concerning the choice of the
Hamiltonian. They also thank Isabelle Catto, Bergfinnur Durhuus and
\'Eric Séré for useful discussions, and acknowledge support from the
European Union's IHP network \emph{Analysis \& Quantum}
HPRN-CT-2002-00277.}

\section{Model and main results}
In this section, we present the models and state our results. The proofs are given separately, in the next section.

\subsection{QED energy on the torus in the Hartree-Fock approximation}\label{questions}
Let us start by defining the no-photon QED Hamiltonian in the cube $\cC_L = [-L/2;L/2)^3$. As mentioned in the introduction, we add periodic boundary conditions and therefore work in the torus $\bT_L:=\R^3/(L\Z^3)$.

\subsubsection*{Notations}
A function $\psi$ in $L^2(\bT_L,\C^4)=L^2_{\rm per}(\cC_L,\C^4)$ can be written as
$$\psi(x)=\frac{(2\pi)^{3/2}}{L^3}\sum_{k\in 2\pi \Z^3/L}\widehat{\psi}(k)e^{ik\cdot x}=\left(\frac{2\pi}{L}\right)^{3/2}\sum_{k\in 2\pi \Z^3/L}\widehat{\psi}(k)e_k(x),$$
where $e_k(x)={e^{ik\cdot x}}/{L^{3/2}}$ and $\widehat{\psi}(k)\in\C^4$.
We now add a cut-off in Fourier space and define the following finite-dimensional subspace of $L^2(\bT_L,\C^4)$
\begin{equation}
\gH_\Lambda^L:={\rm span}\left\{e_k\varepsilon_\sigma\ |\ k\in\Gamma_\Lambda^L,\ \sigma\in\{1,...,4\}\right\},\qquad \Gamma_\Lambda^L:=2\pi \Z^3/L \cap B(0,\Lambda),
\label{cutoff1}
\end{equation}
$(\varepsilon_\sigma)_{\sigma=1,..,4}$ being the canonical orthonormal basis of $\C^4$.
Note that any operator $\gamma$ acting on $\gH_\Lambda^L$ has a kernel of the form
$$\gamma(x,y)=\sum_{k,l\in\Gamma_\Lambda^L}\widehat{\gamma}(k,l)e_k(x)\overline{e_l(y)}$$
where $\widehat{\gamma}(k,l)$ is a $\C^4\times\C^4$ matrix such that $\gamma[\varepsilon_\sigma e_k]=\sum_{\sigma'}\sum_{l\in\Gamma_\Lambda^L}\widehat{\gamma}(l,k)_{\sigma',\sigma}e_l\varepsilon_{\sigma'}$.
Its density is defined as
$$\rho_\gamma(x)=\tr_{\C^4}(\gamma(x,x))=L^{-3}\sum_{k,l\in\Gamma_\Lambda^L}\tr_{\C^4}\left(\widehat{\gamma}(k,l)\right)e^{ix\cdot(k-l)}.$$
A translation-invariant operator $T$ acting on $\gH_\Lambda^L$ satisfies $\widehat{T}(k,l)=g(k)\delta_{kl}$ where $g(k)$ is, for any $k\in\Gamma_\Lambda^L$, a $\C^4\times\C^4$ matrix. Denoting now
$$\check{g}(x)=(2\pi/L)^{3/2}\sum_{k\in\Gamma_\Lambda^L}g(k)e_k(x),$$
 we easily see that
$$T(x,y)=(2\pi)^{-3/2}\check{g}(x-y).$$
The density of a translation-invariant operator is a constant:
$$\rho_{T}(x):=\tr_{\C^4}T(x,x)=(2\pi)^{-3/2}\tr_{\C^4}\check{g}(0)=\frac{1}{L^3}\sum_{k\in\Gamma_\Lambda^L}\tr_{\C^4}g(k).$$
The identity on $\gH_\Lambda^L$ is $I_\Lambda^L$ whose kernel is
\begin{equation}
I_\Lambda^L(x,y)=\frac{1}{L^3}\sum_{k\in\Gamma_\Lambda^L}I_4\,e^{ik\cdot(x-y)}
\label{def_identity}
\end{equation}
and whose density is $\rho_{I_\Lambda^L}=4|\Gamma_\Lambda^L|/L^3$.
In the following, we shall denote
\begin{equation}
\label{def_rho_id}
\rho_\Lambda^L=\frac{\rho_{I_\Lambda^L}}{2}=\frac{2|\Gamma_\Lambda^L|}{L^3}.
\end{equation}

\subsubsection*{Dirac operator, Coulomb potential and external field on $\bT_L$}
Recall that the free Dirac operator is defined as
$$D^0=-i\alp\cdot\nabla+m_0\beta$$
where $m_0>0$ is the bare mass of the electron and $\alp=(\alpha_1,\alpha_2,\alpha_3)$, $\beta$ are the Dirac matrices \cite{Thaller}.
 On the torus, we use the same notation and simply define $D^0$ as the multiplication operator in the Fourier domain, by $(\alp\cdot k+m_0\beta)_{k\in\Gamma_\Lambda^L}$. The same abuse of notation will be done for the operator $-i\nabla$ acting as a multiplication operator in the Fourier domain by $(k)_{k\in\Gamma_\Lambda^L}$.

We shall use a Coulomb potential similar to the choices made in \cite{LS,CLL1,CLL,CP}. We define $W_L$ as being the unique solution of
\begin{equation}
\label{def_Coulomb}
\left\{\begin{array}{l}
\displaystyle-\Delta W_L=4\pi\left(\sum_{{\rm x}\in\Z^3}\delta_{L\rm x}-\frac{1}{L^3}\right)\\
\displaystyle\min_{\cC_L} W_L=0.
\end{array} \right.
\end{equation}
This means that there exists a constant $\mu>0$ such that
$$W_L(x)=\frac{1}{L^3}\left(\sum_{\substack{k\in(2\pi)\Z^3/L\\ k\neq0}}\frac{4\pi}{|k|^2}e^{ik\cdot x}+\mu L^2\right).$$
Notice that in \cite{LS,CLL1,CLL,CP}, $\mu$ is replaced by 0 (i.e. the integral of the Coulomb potential is assumed to vanish). Indeed most of our results are valid if $\mu$ is replaced by any non-negative real number. However, the non-negativity of the Coulomb potential, which seems physically relevant to us,  is usually needed in the study of Hartree-Fock minimizers. Moreover, our choice better mimics the behavior of $1/|k|^2$ at 0.

In the following, we shall also consider an external potential of the form
$$\phi=\alpha n\ast\frac{1}{|\cdot|}$$
in the whole space, created by external sources for instance like nuclei. We will assume that $n\in \mathcal{C}$ where $\mathcal{C}$ is the so-called Coulomb space \cite{HLS2}
\begin{equation}
\mathcal{C}=\left\{f\ |\ \int_{\R^3}\frac{|\widehat{f}(k)|^2}{|k|^2}dk<\ii\right\}.
\label{def_C}
\end{equation}
We do not necessarily assume in this paper that $n$ is an $L^1$ non-negative function but the reader should think of $\int_{\R^3}n=Z$ as being the total number of protons in the nuclei.

\begin{remark}
Notice that, since we will add an ultraviolet cut-off, the definition \eqref{def_C} also contains the usual Coulomb potential since the (regularized) Dirac measure $\delta_\Lambda$ defined by
\begin{equation}
\widehat{\delta_\Lambda}=(2\pi)^{-3/2}\1_{B(0,\Lambda)}
\label{def_delta}
\end{equation}
 belongs to the space $\cC$.
\end{remark}

For the sake of simplicity, we shall also assume that the restriction to the ball $B(0,\Lambda)$ of the Fourier transform $\widehat{n}$ is a continuous function. This allows us to define the external potential on the torus by
\begin{equation}
n_L(x)=\frac{(2\pi)^{3/2}}{L^3}\sum_{k\in\Gamma^L_\Lambda}\widehat{n}(k)e^{ik\cdot x},
\label{def_n_L}
\end{equation}
In the whole space, the Dirac operator with external potential will be denoted by
$D^\phi=D^0-\phi=D^0-\alpha n\ast\frac{1}{|\cdot|},$
whereas we use the notation
\begin{equation}
D^\phi_L=D^0-\phi_L,\qquad \phi_L(x):=\alpha n_L\ast W_L(x)=\alpha \int_{\cC_L}n_L(y) W_L(x-y)dy
\label{def_Dirac_Coulomb2}
\end{equation}
for the corresponding operator acting on $\gH_\Lambda^L$.

\subsubsection*{No-photon QED Hamiltonian on $\bT_L$}
We are now able to define and compute the no-photon QED energy in $\bT_L$.
The Fock space associated with $\gH_\Lambda^L$ is
$$\F_\Lambda^L:=\C\oplus\bigoplus_{m\geq1}\left(\bigwedge_{i=1}^m\gH_\Lambda^L\right)$$
(this is indeed a finite dimensional space).
On $\F_\Lambda^L$, the creation operator $\Psi^*_{k,\sigma}$ is defined as usually, for $k\in\Gamma_\Lambda^L$ and $\sigma\in\{1,2,3,4\}$, by
$$\Psi^*_{k,\sigma}(\psi_1\wedge\cdots\wedge \psi_p)=(e_k\varepsilon_\sigma)\wedge \psi_1\wedge\cdots\wedge \psi_p.$$
It satisfies the anti-commutation relation
\begin{equation}
\{\Psi_{k,\sigma},\Psi^*_{l,\sigma'}\}=\delta_{k,l}\delta_{\sigma,\sigma'}.
\label{CAR}
\end{equation}
In the following, we shall use the common notation $\Psi(x)_\sigma=\sum_{k\in\Gamma_\Lambda^L}e_k(x)\Psi_{k,\sigma}$ for the second-quantized field operator. We shall also need to use, for some $f\in\gH_\Lambda^L$,  the notation
\begin{equation}
\label{boldPsi}
\bPsi(f)_\sigma=\left(\frac{2\pi}{L}\right)^{3/2}\sum_{k\in\Gamma_\Lambda^L}\widehat{f}(k)_\sigma\Psi_{k,\sigma}
\end{equation}
when $f=\left(\frac{2\pi}{L}\right)^{3/2}\sum_{k\in\Gamma_\Lambda^L}\widehat{f}(k)e_k$.

The no-photon QED Hamiltonian is the well-defined operator acting on the finite-dimensional space $\F_\Lambda^L$,
\begin{multline}
\label{Hamiltonian}
\mathbb{H}^\phi_L = \int_{\bT_L} \Psi^*(x) D^0 \Psi(x) \,dx -\int_{\bT_L}\phi_L(x)\rho(x)dx\\+ \frac{\alpha}2 \iint_{(\bT_L)^2} \rho(x)\rho(y)W_L(x-y)dx\,dy
\end{multline}
where $\alpha$ is the \emph{bare} fine structure constant and
\begin{eqnarray}
\rho(x) & = & \frac{1}{2}\tr[\Psi^*,\Psi](x)\nonumber\\
 & = & \frac{1}{2}\sum_{\sigma}\left(\Psi^*(x)_\sigma\Psi(x)_\sigma-\Psi(x)_\sigma\Psi^*(x)_\sigma\right)\nonumber\\
 & = & \frac{1}{2}\sum_{\sigma}\sum_{k,l\in\Gamma_\Lambda^L}\left(\Psi^*_{k,\sigma}\Psi_{l,\sigma}-\Psi_{l,\sigma}\Psi^*_{k,\sigma}\right)e_k(x)\overline{e_l(x)}.\label{formula_rho_CAR}
\end{eqnarray}
The charge-conjugation operator $\CJ$ is the uniquely defined (up to a phase) unitary operator acting on the Fock space $\F_\Lambda^L$ (see, e.g. \cite[Prop. $2.1$]{Scharf}) such that, for any $f\in\gH_\Lambda^L$,
$$\CJ\bPsi(f)\CJ^{-1}=\bPsi(C f)^*\ \text{ and }\  \CJ\bPsi(f)^*\CJ^{-1}=\bPsi(C f),$$
where $C$ is the charge conjugation operator acting on $\gH_\Lambda^L$, defined by $C f:=i\beta\alpha_2\overline{f}$, and $\bPsi(f)$ has been introduced in \eqref{boldPsi}. It is then easy to see that the following relations hold
$$\CJ\rho(x)\CJ^{-1}=-\rho(x),\qquad \CJ\mathbb{H}^\phi_L \CJ^{-1}=\mathbb{H}^{-\phi}_L.$$
Using the CAR \eqref{CAR} and the formula \eqref{formula_rho_CAR}, we obtain
\begin{equation}
\rho(x)=\Psi^*(x)\Psi(x)-\rho_\Lambda^L.
\label{formula_rho}
\end{equation}
where $\rho_\Lambda^L$ is defined in \eqref{def_rho_id}.
Inserting \eqref{formula_rho}, we obtain for the last term of \eqref{Hamiltonian}
\begin{align}
& \frac{\alpha}{2}\iint_{(\bT_L)^2} \rho(x)\rho(y)W_L(x-y)dx\,dy\nonumber\\
&=   \frac{\alpha}{2}\iint_{(\bT_L)^2} \Psi^*(x)\Psi(x)\Psi^*(y)\Psi(y)W_L(x-y)dx\,dy\nonumber\\
& \quad -\alpha\rho^L_\Lambda\iint_{(\bT_L)^2}\Psi^*(x)\Psi(x)W_L(x-y)dx\,dy +\frac{\alpha(\rho^L_\Lambda)^2}{2}\iint_{(\bT_L)^2}W_L(x-y)dx\,dy\nonumber\\
  &=  \alpha\sum_{1\leq i<j}W_L(x_i-x_j)+\frac\alpha2\iint_{(\bT_L)^2}I_\Lambda^L(x,y)\Psi^*(x)\Psi(y)W_L(x-y)dx\,dy\nonumber\\
&\quad  -\alpha\rho_\Lambda^L\iint_{(\bT_L)^2}\Psi^*(x)\Psi(x)W_L(x-y)dx\,dy+\frac{\alpha(\rho^L_\Lambda)^2}{2}\iint_{(\bT_L)^2}W_L(x-y)dx\,dy,\label{formula_Ham_WL}
\end{align}
where we recall that $I_\Lambda(x,y)$ is defined in \eqref{def_identity}.
Therefore, computing the energy of a Hartree-Fock state with density matrix $P(x,y)$ (an orthonormal projector on $\gH_\Lambda^L$), we obtain from \eqref{Hamiltonian} and \eqref{formula_Ham_WL}
\begin{multline}
\label{energy_projector}\E^{\rm\small QED}_{L,\phi}(P) = \tr(D^0P)-\int_{\bT_L}\phi_L(x)(\rho_P(x)-\rho_\Lambda^L)\, dx\\
 +\frac\alpha2\iint_{(\bT_L)^2}\rho_P(x)\rho_P(y)W_L(x-y)dx\,dy-\frac\alpha2\iint_{(\bT_L)^2}|P(x,y)|^2W_L(x-y)dx\,dy  \\
+\frac\alpha2\int_{(\bT_L)^2}\tr_{\C^4}\left(I_\Lambda^L(x,y)P(x,y)\right)W_L(x-y)dx\,dy\\
 -\alpha\rho_\Lambda^L\iint_{(\bT_L)^2}\rho_P(x)W_L(x-y)dx\,dy +\frac{\alpha(\rho^L_\Lambda)^2}{2}\iint_{(\bT_L)^2}W_L(x-y)dx\,dy.
\end{multline}
Using $\rho_{I_\Lambda^L/2}(x)=\rho_\Lambda^L$ and $\tr(D^0I^L_\Lambda)=0$, we infer
\begin{equation}
\fbox{$\displaystyle
\E^{\rm\small QED}_{L,\phi}(P)=\E_L^\phi(P-I_\Lambda^L/2)+\frac{\alpha}{8}\iint_{(\bT_L)^2}|I_\Lambda^L(x,y)|^2W_L(x-y)dx\,dy
$}
\label{QED_energy}
\end{equation}
where $\E_L^\phi$ is the usual Dirac-Fock energy on the torus
\begin{equation}
\fbox{$\displaystyle
\E_L^\phi(\gamma)=\tr(D^\phi_L\gamma)+\frac\alpha2 D_L(\rho_\gamma,\rho_\gamma)-\frac\alpha2 \iint_{(\bT_L)^2}|\gamma(x,y)|^2W_L(x-y)dx\,dy
$}
\label{QED_DF}
\end{equation}
defined for any self-adjoint operator $\gamma$ acting on $\gH_\Lambda^L$, and with
$$D_L(f,f)=\iint_{(\bT_L)^2}f(x)f(y)W_L(x-y)dx\,dy=\frac{1}{L^3}\sum_{k\in(2\pi\Z^3)/L}|\widehat{f}(k)|^2\widehat{W_L}(k)\geq0,$$
$$\rho_\gamma(x)=\tr_{\C^4}\gamma(x,x).$$
The last term of \eqref{QED_energy} is a constant which behaves like
\begin{equation}
\frac{\alpha}{8}\iint_{(\bT_L)^2}|I_\Lambda^L(x,y)|^2W_L(x-y)dx\,dy\sim_{L\to\ii}\frac{\alpha L^3}{2}\int_{\R^3}\frac{|\delta_\Lambda(x)|^2}{|x|}dx
\label{add_constant}
\end{equation}
where $\delta_\Lambda\in L^2(\R^3)\cap L^\ii(\R^3)$ is the Fourier inverse of $(2\pi)^{-3/2}\1_{B(0,\Lambda)}$ already defined in \eqref{def_delta}. Because of \eqref{add_constant}, this term shifts the limit of the energy per unit volume by a constant and disappears when looking at differences. Hence, it will not play any role for the study of our thermodynamic limit, and we discard it for the rest of the paper. We therefore study the minimization problem
\begin{equation}
\fbox{$\displaystyle
E_L(\phi):=\inf\left\{\E_L^\phi(\gamma),\ \gamma\in\mathcal{L}(\gH^L_\Lambda),\ \gamma^*=\gamma,\ -\frac{I^L_\Lambda}{2}\leq \gamma \leq \frac{I^L_\Lambda}{2}\right\}$}
\label{min_box}
\end{equation}
($\mathcal{L}(\gH_\Lambda^L)$ denotes the space of all linear operators acting on $\gH_\Lambda^L$). Notice that we have extended the set $\{P-I^L_\Lambda/2,\ P \text{ orth. projector}\}$ to its convex hull
$$\cG_\Lambda^L=\left\{\gamma\in\mathcal{L}(\gH^L_\Lambda),\ \gamma^*=\gamma,\ -\frac{I^L_\Lambda}{2}\leq \gamma \leq \frac{I^L_\Lambda}{2}\right\}$$
as this is usually done in Hartree-Fock type theories \cite{Lieb}. It will be proved that this does not change the minimizer.
With this change of variable, our goal will be to prove
\begin{itemize}
\item {\it without external field ($\phi=0$):}
\begin{enumerate}
\item the existence of a unique minimizer $\gamma_L^0=\cP_L^0-I_\Lambda^L/2$ for $E_L(0)$, when $L$ is large enough;
\item that the energy per unit volume $\displaystyle\frac{E_L(0)}{L^3}$ has a limit as $L\to\ii$;
\item that $\gamma^0_L$ converge in a certain sense to a translation-invariant limit $\gamma^0=\cP^0_--1/2$, which will be the HF free vacuum ($\cP^0_-$ is a projector);
\item that the limiting $\gamma^0$ is the unique minimizer of the \emph{energy per unit volume}, a functional defined only for translation-invariant operators acting on the whole space;
\end{enumerate}
\item {\it in the presence of an external field ($\phi\neq0$):}
\begin{enumerate}\addtocounter{enumi}{4}
\item the existence of a minimizer $\gamma_L=\cP_L-I_\Lambda^L/2$ for $E_L(\phi)$;
\item that the energy difference $E_L(\phi)-E_L(0)$ has a limit as $L\to\ii$;
\item that $\gamma_L$ converges in a certain sense to an operator $\bar\gamma=\bar\cP_--1/2$ which will be interpreted as the HF polarized vacuum ($\bar\cP_-$ is a projector);
\item that $\bar\gamma-\gamma^0$ is a minimizer of the Bogoliubov-Dirac-Fock energy, which measures the energy with respect to the (infinite) energy of the free vacuum $\gamma^0$.
\end{enumerate}
\end{itemize}

These questions are rather common in the mathematical study of thermodynamic limits \cite{LL,LS,CLL,CLL1}.
Before we answer them, we have to define properly the variational problems obtained in the whole space. In the next  section, we define the energy per unit volume of translation-invariant operators acting on the whole space when no external field is present; its minimizer will be the free vacuum $\gamma^0$. In Section \ref{sec_BDF}, we define the associated Bogoliubov-Dirac-Fock model properly and prove the existence of the polarized vacuum $\bar \gamma$ in the presence of an external field; this will be only an easy extension of the work already done by Hainzl-Lewin-Séré in \cite{HLS1,HLS2}. Finally, we answer questions (1)--(8) in Section \ref{sec_thermo}.

\subsection{Definition of the free vacuum $\cP^0_-$}\label{sec_free_vac}
Let us now define the models in the whole space. The cut-off is implemented in the Fourier domain by considering the following Hilbert space
\begin{equation}
\gH_\Lambda:=\left\{\psi \in L^2(\R^3, \C^4)\ |\ {\rm supp}\, \hat \psi \subset B(0,\Lambda)\right\}.
\label{cutoff2}
\end{equation}
In this subsection, we consider the case where no external field is present, $n=0$. We want to define the energy per unit volume of a translation-invariant operator $\gamma$ acting on $\gH_\Lambda$ and such that
$$-\frac{I_\Lambda}{2}\leq\gamma\leq \frac{I_\Lambda}{2},$$
$I_\Lambda$ being the identity on $\gH_\Lambda$. Such an operator acts as a multiplication operator in the Fourier domain and can therefore be written $\gamma={f}(-i\nabla)$ or, formally $\gamma(x,y)=(2\pi)^{-3/2}\check{f}(x-y)$, with $f$ belonging to
$$\A_\Lambda=\left\{f\in L^\ii(B(0,\Lambda),\S_4(\C)),\ -\frac{I_{\C^4}}{2}\leq f \leq \frac{I_{\C^4}}{2}\right\}.$$
Here $\S_4(\C)$ is the set of $4\times4$ self-adjoint complex matrices. Notice that its density of charge is a well-defined constant:
$$\rho_{\gamma}(x)=\tr_{\C^4}\gamma(x,x)=(2\pi)^{-3/2}\tr_{\C^4}\check{f}(0)=(2\pi)^{-3}\int_{B(0,\Lambda)}\tr_{\C^4}(f(p))\,dp.$$
The energy per unit volume of such a translation-invariant operator $\gamma=f(-i\nabla)$ is then defined in terms of $f\in\A_\Lambda$ by
\begin{equation}
\fbox{$\displaystyle
\T(f)=\frac{1}{(2\pi)^{3}}\int_{B(0,\Lambda)}\tr_{\C^4}[D^0(p)f(p)]dp-\frac{\alpha}{(2\pi)^5}\iint_{B(0,\Lambda)^2}\frac{\tr_{\C^4}[f(p)f(q)]}{|p-q|^2}dp\,dq.$}
\label{def_fn_F}
\end{equation}

To give a hint why we consider the energy $\T$, let us fix some operator $\gamma=f(-i\nabla)$ where $f\in\A_\Lambda$. To simplify the discussion, we also assume that $\tr_{\C^4}(f(k))=0$ for any $k\in B(0,\Lambda)$. To avoid any confusion, we denote by $\gamma^L$ the operator acting on $\gH_\Lambda^L$ associated with $\gamma$, and which is just the multiplication operator by $(f(k))_{k\in\Gamma_\Lambda^L}$ in the Fourier domain.
Its energy is given by \eqref{QED_DF} and can be expressed as
$$
\E_L^0(\gamma^L) = \sum_{k\in\Gamma_\Lambda^L}\tr_{\C^4}(D^0(k)f(k))-L^3\frac{\alpha}{2(2\pi)^3}\int_{\bT_L}|\check{f_L}(x)|^2W_L(x)dx
$$
where $\check{f_L}(x):=\left(2\pi/L\right)^{3/2}\sum_{k\in\Gamma_\Lambda^L}{f}(k)e_k(x)$. Notice that we have used
$$\rho_{\gamma^L}=L^{-3}\sum_{k\in\Gamma_\Lambda^L}\tr_{\C^4}f(k)=0$$
 since by assumption $\tr_{\C^4}f(k)=0$ for any $k\in B(0,\Lambda)$.
It is easy to see that
$$\lim_{L\to\ii}\frac{\E_L^0(\gamma^L)}{L^{3}}=\T(f).$$
It is therefore natural to define the free vacuum as a minimizer of $\T$ and we introduce
$$E^\T:=\inf\left\{\T(f)\ |\ f\in\A_\Lambda\right\}.$$

\begin{theorem}[Definition of the free vacuum]\label{free_vac}
Assume that $0\leq\alpha<4/\pi$, $\Lambda>0$ and that $m_0>0$. Then $\T$ possesses a unique global minimizer $\bar f$ on $\A_\Lambda$ and any minimizing sequence $(f_n)\subset\A_\Lambda$ of $\T$ converges strongly to $\bar f$ in $L^2(B(0,\Lambda),\S_4(\C))$.

The associated renormalized density matrix $\gamma^0:=\bar f(-i\nabla)$ is a translation-invariant operator satisfying the self-consistent equation
\begin{equation}
\left\{\begin{array}{l}
\displaystyle\gamma^0=-\frac{\sgn(\cD^0)}{2},\\
\displaystyle\cD^0=D^0-\alpha\frac{\gamma^0(x,y)}{|x-y|}
\end{array} \right.
\label{scf_proj}
\end{equation}
or, written in terms of the translation-invariant projector $\cP^0_-=\gamma^0+I_\Lambda/2$,
\begin{equation}
\cP^0_-=\chi_{(-\ii;0)}\left(\cD^0\right).
\label{scf_proj2}
\end{equation}
Moreover, $\cD^0$ takes the special form, in the Fourier domain,
\begin{equation}
\cD^0(p)=\alp\cdot \omega_p g_1(|p|)+g_0(|p|)\beta
\label{form_new_D}
\end{equation}
where $\omega_p=p/|p|$ and $g_0$, $g_1\in L^\ii([0;\Lambda],\R)$ are such that
\begin{eqnarray}
 &  g_1(x)\geq x \text{ and } g_0(x)\geq m_0,\label{propD1}\\
 &  \displaystyle m_0g_1(x)\leq g_0(x)x\label{propD2}
\end{eqnarray}
for any $x\in[0;\Lambda)$, and therefore
\begin{equation}
m_0^2+|p|^2\leq |\cD^0(p)|^2\leq \frac{g_0(|p|)}{m_0}(m_0^2+|p|^2).
\label{propD3}
\end{equation}
Finally, $\cD^0(p)\in\bigcap_{m\geq1}H^m(B(0,\Lambda))\subset\mathcal{C}^\ii(\overline{B(0,\Lambda)})$.
\end{theorem}

The proof of Theorem \ref{free_vac} is given in section \ref{proof_free_vac}.

This result is a generalization of a work by Lieb and Siedentop. In \cite{LSie}, the equation \eqref{scf_proj} is solved  by means of a fixed point method, only valid under a condition of the form $\alpha\log\Lambda\leq C$. Thanks to the variational interpretation using the function $\T$, we have been able to prove the existence of a solution of \eqref{scf_proj} without any constraint linking $\alpha$ and $\Lambda$, and by means of a completely different proof. Our solution coincides with \cite{LSie} when $\alpha\log\Lambda\leq C$, and the properties of $\cD^0$ stated in Theorem \ref{free_vac} are exactly the ones which have been proved in \cite{LSie}.
The interpretation of \eqref{scf_proj} given in \cite{LSie} however does not seem to be the same as ours.

We notice that a self-consistent equation similar to \eqref{scf_proj} (written in terms of the four-dimensional Green function of the electron) has been approximately solved first by Landau {\it et al.} in \cite{Lan81,Lan89}, and then by many authors (see, e.g., \cite{mass1,mass2,mass3}). They use an ansatz analogous to \eqref{form_new_D} (see, e.g., \cite[Eq. $(1)$]{Lan81} and \cite[Eq. $(4.1)$]{Lan89}).

Remark that the density of the vacuum $\gamma^0=\bar f(-i\nabla)$ also vanishes, since \eqref{scf_proj} and \eqref{form_new_D} mean
$$\bar f(p)=-\frac{g_1(|p|)}{2\sqrt{g_1(|p|)^2+g_0(|p|)^2}}\alp\cdot \omega_p-\frac{g_0(|p|)}{2\sqrt{g_1(|p|)^2+g_0(|p|)^2}}\beta$$
and therefore
$$\rho_{\gamma^0}=(2\pi)^{-3}\int_{B(0,\Lambda)}\tr_{\C^4}\bar f(p)\, dp=0,$$
the Dirac matrices $\alpha_i$ and $\beta$ being trace-less.

\begin{remark} Adapting the proof given in Section \ref{proof_free_vac}, it can be shown that when $m_0=0$, the functional $\T$ also has a unique minimizer $\bar f(p)=-\cD^0(p)/(2|\cD^0(p)|)$, with $\cD^0(p)=\alp\cdot\omega_p g_1(|p|)$, $g_1$ being given by \cite[Eq. $(26)$]{LSie}.
\end{remark}

\begin{remark} The energy $\T$ satisfies a nice scaling invariance property (already used in \cite{CIL}). Namely one has, with an obvious notation,
$$\T_{m_0,\alpha,\Lambda}(f)=\lambda^{-4}\T_{\lambda m_0,\alpha,\lambda\Lambda}(f(\cdot/\lambda))$$
which implies that
$$E^\T(m_0,\alpha,\Lambda)=\lambda^{-4}E^\T(\lambda m_0,\alpha,\lambda\Lambda)\ \text{ and }\ (\cP^0_-)_{\lambda m_0,\alpha,\lambda\Lambda}(p)=(\cP^0_-)_{m_0,\alpha,\Lambda}(p/\lambda).$$
\end{remark}

\subsection{The Bogoliubov-Dirac-Fock theory based on $\cP^0_-$}\label{sec_BDF}
In the previous section, we have defined the HF free vacuum, whose renormalized density matrix is $\gamma^0=\cP^0_--I_\Lambda/2$. We now assume that $n\neq 0$ (recall that $\phi=\alpha n\ast 1/|\cdot|$) and define the Bogoliubov-Dirac-Fock model \cite{CI,CIL,Chaix,BBHS,HLS1,HLS2}, based on this new reference. Our goal will be to show that the results by Hainzl-Lewin-Séré \cite{HLS1,HLS2} can be extended to this case. We refer the reader to \cite{CI,Chaix,HLS1,HLS2} for a detailed presentation of the BDF model.

The BDF energy reads \cite{HLS1,HLS2}
\begin{equation}
\label{BDF1}
\fbox{$\displaystyle
\E^\phi_{\rm BDF}(Q):= \tr_{\cP^0_-}(\cD^0Q)+\frac{\alpha}{2}D(\rho_Q,\rho_Q)-\alpha D(\rho_Q,n)-\frac{\alpha}{2}\iint_{\R^6}\frac{|Q(x,y)|^2}{|x-y|}dx\,dy$}
\end{equation}
where
\begin{equation}
Q\in  \mathcal{Q}_{\Lambda}:=\left\{Q\in\gS^{\cP^0_-}_1(\gH_\Lambda),\ -\cP^0_-\leq Q\leq \cP^0_+,\ \rho_Q\in\cC\right\},
\label{def_Q_Lambda}
\end{equation}
\begin{equation*}
\mathcal{C}=\left\{f\ |\ D(f,f)<\ii\right\}
\label{def_C2}
\end{equation*}
and
$$D(f,g)=4\pi\int\frac{\overline{\widehat{f}(k)}\widehat{g}(k)}{|k|^2}dk.$$
The space $\gS^{\cP^0_-}_1(\gH_\Lambda)$ is introduced in \cite[Sec. 2.1]{HLS1}. It contains all the Hilbert-Schmidt operators $Q\in\gS_2(\gH_\Lambda)$ which are such that $Q^{++}=\cP^0_+Q\cP^0_+$ and $Q^{--}=\cP^0_-Q\cP^0_-$ are trace-class ($\in\gS_1(\gH_\Lambda)$). The $\cP^0_-$-trace of $Q$ is then defined by
$$\tr_{\cP^0_-}Q:=\tr(Q^{++}+Q^{--}).$$
Finally, $\rho_Q(x)=\tr_{\C^4}Q(x,x)$ is the charge density associated with $Q$, a well-defined object in Fourier space thanks to the ultraviolet cut-off \cite{HLS1}:
$$\widehat{\rho_Q}(k)=(2\pi)^{-3/2}\int_{B(0,\Lambda)}\tr_{\C^4}\left(\widehat{Q}(p+k/2,p-k/2)\right)\, dp.$$

Notice that compared to \cite{CI,CIL,Chaix,BBHS,HLS1,HLS2}, we have not only replaced $P^0_-$ by $\cP^0_-$, but also $D^0$ by $\cD^0$ in the definition \eqref{BDF1} of the BDF energy.

The interpretation of \eqref{BDF1} is that $\E_{\rm BDF}^\phi(\gamma-\gamma^0)$ is the energy of the HF state $\gamma$, measured relatively to the energy of the free vacuum $\gamma^0$.
This statement will be made more precise in the next section, in which we prove that $E_L(\phi)-E_L(0)$ converges to the minimum of the BDF energy. Notice when
$\gamma$ satisfies $-I_\Lambda/2\leq\gamma\leq I_\Lambda/2$, then one has $-\cP^0_-\leq\gamma-\gamma^0\leq\cP^0_+$, justifying the constraint imposed on $Q$ in the definition of $\mathcal{Q}_\Lambda$ in \eqref{def_Q_Lambda}.

We may now state a result similar to \cite[Theorem 1]{HLS1}:
\begin{theorem}[The BDF energy is bounded-below]\label{BDF_bound_below} Assume that $0\leq\alpha\leq4/\pi$, $\Lambda>0$, $m_0>0$ and that $n\in\cC$.
\begin{enumerate}
\item One has
$$\forall Q\in\mathcal{Q}_\Lambda,\qquad \E_{\rm BDF}^\phi(Q)+\frac\alpha2 D(n,n)\geq0$$
and therefore $\E$ is bounded from below on $\mathcal{Q}_\Lambda$.
\item If moreover $n=0$, then $\E_{\rm BDF}^0$ is non-negative on $\mathcal{Q}_\Lambda$, $0$ being its unique minimizer.
\end{enumerate}
\end{theorem}

\begin{proof}
The proof is the same as in \cite{CIL,BBHS,HLS1}, using Kato's inequality
$$\frac{1}{|x|}\leq \frac{\pi}{2}|D^0|\leq \frac{\pi}{2}|\cD^0|,$$
due to  \eqref{propD3}.
\end{proof}

The interpretation of the second part of Theorem \ref{BDF_bound_below} is that $\gamma^0$ is not only a minimizer among translation-invariant projectors (Section \ref{sec_free_vac}), but also among Hilbert-Schmidt perturbations.

Let us now define the BDF ground state energy in the presence of an external field:
\begin{equation}
E_{\rm BDF}(\phi)=\inf_{\mathcal{Q}_\Lambda}\E^\phi_{\rm BDF}.
\end{equation}
The existence of a minimizer is obtained by a result analogous to \cite[Theorem 1]{HLS2}:
\begin{theorem}[Definition of the polarized vacuum]\label{BDF_min} Assume that $0\leq\alpha<4/\pi$, $\Lambda>0$, $m_0>0$ and that $n\in\cC$. Then $\E_{\rm BDF}^\phi$ possesses a minimizer $\bar Q$ on $\mathcal{Q}_\Lambda$ such that $\bar\gamma=\bar Q+\gamma^0$ satisfies the self-consistent equation
\begin{equation}
\label{eq_gamma}
\left\{\begin{array}{l}
\displaystyle\bar\gamma=-\frac{\sgn(\bar\cD)}{2},\\
\displaystyle\bar\cD:= D^0+\alpha\left(\rho_{\bar\gamma}-n\right)\ast\frac{1}{|\cdot|}-\alpha\frac{\bar\gamma(x,y)}{|x-y|}
\end{array} \right.
\end{equation}
or, written in terms of the projector $\bar\cP_-:=\bar\gamma+I_\Lambda/2=\bar Q+\cP^0_-$,
\begin{equation}
\bar\cP_- =\chi_{(-\ii;0)}\left(\bar\cD\right).
\end{equation}
Additionally, if $\alpha$ and $n$ satisfy
\begin{equation}
0\leq\alpha\frac\pi4\left\{1-\alpha\left(\frac\pi2 \sqrt{\frac{\alpha/2}{1-\alpha\pi/4}}+\pi^{1/6}2^{11/6}\right)\|n\|_{\mathcal{C}}\right\}^{-1}\leq1,
\label{condition_uniqueness}
\end{equation}
then this global minimizer $\bar Q$ is unique and the associated polarized vacuum is neutral:
$\tr_{\cP^0_-}(\bar Q)=0$.
\end{theorem}
\begin{proof}
The proof works analogously to \cite[Proof of Theorem 1]{HLS2}. The only modification is \cite[Lemma 1]{HLS2}: one proves that $\norm{\,[\xi_R,|\cD^0|]\,}_{\gS_\ii(\gH_\Lambda)}=O\left(1/R\right)$ using the regularity of $\cD^0(p)$ stated in Theorem \ref{free_vac}.
\end{proof}

\subsection{Thermodynamic limits}\label{sec_thermo}
Let us first state a result for the thermodynamic limit when $n=0$, and which answers to the questions (1)--(4) of Section \ref{questions}. We recall that $E^\T$, $\T$, $\bar f$, $\gamma^0$ and $\cP^0_-$ are defined in Section \ref{sec_free_vac}.
\begin{theorem}[Thermodynamic limit with no external field]\label{thermo1} Assume that $0\leq\alpha<4/\pi$, $\Lambda>0$, and that $m_0>0$. Then for $L$ large enough, $\E_L^0$ possesses a unique minimizer $\gamma_L^0=\cP^0_L-I_\Lambda^L/2$ on $\cG_\Lambda^L$, where $\cP^0_L$ is an orthogonal projector. It is translation-invariant, $\gamma_L^0=f_L^0(-i\nabla)$ (or equivalently $\gamma_L^0(x,y)=(2\pi)^{-3/2}\check{f}^0_L(x-y)$). One has
\begin{equation}
\lim_{L\to\ii}\frac{E_L(0)}{L^3} = E^\T= \min\left\{\T(f)\ |\ f\in\A_\Lambda\right\}
\label{limit_thermo}
\end{equation}
and
\begin{equation}
\lim_{L\to\ii}\norm{\gamma^0_L-\gamma^0}_{\gS_\ii(\bT_L)}=\lim_{L\to\ii}\sup_{k\in\Gamma_\Lambda^L}\left|f^0_L(k)-\bar f(k)\right|=0,
\label{limit_sup0}
\end{equation}
\begin{equation}
\lim_{L\to\ii}\norm{\cP^0_L-\cP^0_-}_{\gS_\ii(\bT_L)}=\lim_{L\to\ii}\sup_{k\in\Gamma_\Lambda^L}\left|\cP^0_L(k)-\cP^0_-(k)\right|=0.
\label{limit_sup}
\end{equation}
\end{theorem}

The proof of this result is given in Section \ref{proof_thermo1}.


It is interesting to see how this result would change if another choice was made for the definition of the no-photon QED Hamiltonian on the torus, like
$$\tilde{\mathbb H}^1_L:=\int_{\bT_L} \Psi^*(x) D^0 \Psi(x) \,dx + \frac{\alpha}2 \iint_{(\bT_L)^2} \Psi^*(x)\Psi(x)\Psi^*(y)\Psi(y)W_L(x-y)\,dx\,dy$$
or
\begin{align*}
\tilde{\mathbb H}_L^2 &:=\sum_{i\geq1} (D^0)_{x_i}+\sum_{1\leq i<j}W_L(x_i-x_j)\\
&=\int_{\bT_L} \Psi^*(x) D^\phi_L \Psi(x) \,dx + \frac{\alpha}2 \iint_{(\bT_L)^2} \Psi^*(x)\Psi^*(y)\Psi(y)\Psi(x)W_L(x-y)\,dx\,dy.
\end{align*}
In both cases, the thermodynamic limit is dramatically changed since we can prove the following result:
\begin{theorem}[Thermodynamic limit for other Hamiltonians]\label{thermo1b} Assume that $\mathbb{H}_L^0$ is replaced by $\tilde{\mathbb H}_L^1$ or $\tilde{\mathbb H}_L^2$ in the previous study, that $\alpha>0$, $\Lambda>0$, $m_0> 0$ and $n=0$. Let us denote by $0\leq\Theta_L\leq I_\Lambda^L$ a self-adjoint operator acting on $\gH_\Lambda^L$, which is the one-body density matrix of a minimizer of the QED Hamiltonian, when it is restricted to translation-invariant Hartree-Fock states. Then
\begin{equation}
\lim_{L\to\ii}\rho_{\Theta_L}=\lim_{L\to\ii}L^{-3}\sum_{k\in\Gamma_\Lambda^L}\tr_{\C^4}(\Theta_L(k))=0.
\label{limit_other}
\end{equation}
Therefore, writing $\Theta_L(x,y)=(2\pi)^{-3/2}{\xi}_L(x-y)$,
\begin{equation}
\lim_{L\to\ii}\norm{\xi_L}_{L^\ii(\R^3)}=0.
\label{limit_pL}
\end{equation}
\end{theorem}

In other words, when the Hamiltonian $\tilde{\mathbb H}^1_L$ or $\tilde{\mathbb H}^2_L$ is chosen instead of $\mathbb{H}^0_L$, the thermodynamic limit as $L\to\ii$ is trivial. This phenomenon is due to the term $D_L(\rho,\rho)$ which behaves like $L^5$ and not like $L^3$ and which plays the role of a \emph{penalization}: the limit necessarily has a vanishing density. This shows the usefulness of the commutator in the definition of $\rho(x)$ in \eqref{formula_rho_CAR} or \eqref{Rho1}. In that case, one obtains
$$\lim_{L\to\ii}\rho_{\gamma_L^0}=\lim_{L\to\ii}\rho_{(\cP^0_L-I_\Lambda^L/2)}=\lim_{L\to\ii}(\rho_{\cP^0_L}-\rho_\Lambda^L/2)=0$$
and $\rho_{\gamma^0}=\rho_{(\cP^0_--\cP^0_+)/2}=0$.

\medskip

We now state a result for the thermodynamic limit when $n\neq0$; this answers the questions (5)--(8) of Section \ref{questions}.
\begin{theorem}[Thermodynamic limit with external field]\label{thermo2} Assume that $0\leq\alpha<4/\pi$, $\Lambda>0$, $m_0>0$, $n\in\mathcal{C}$ and that $\widehat{n}$ is continuous on $B(0,\Lambda)$. Then for any $L$, $\E_L^\phi$ possesses a minimizer $\gamma_L=\cP_L-I_\Lambda^L/2$ on $\cG_\Lambda^L$ where $\cP_L$ is an orthogonal projector, and one has
\begin{equation}
\lim_{L\to\ii}\left\{E_L(\phi)-E_L(0)\right\} =E_{\rm BDF}(\phi) =\min\left\{\E_{\rm BDF}^\phi(Q),\ Q\in\mathcal{Q}_\Lambda\right\}.
\label{limit_thermo2}
\end{equation}
Moreover, up to a subsequence, $Q_L(x,y):=(\gamma_L-\gamma_L^0)(x,y)=(\cP_L-\cP^0_L)(x,y)$ converges uniformly on compact subsets of $\R^6$ to $\bar Q(x,y)$, a minimizer of $\E_{\rm BDF}^\phi$ on $\mathcal{Q}_\Lambda$.
\end{theorem}

This result justifies the Bogoliubov-Dirac-Fock theory for the study of the polarized vacuum in no-photon QED restricted to Hartree-Fock states. The proof is given in Section \ref{proof_thermo2}.


\bigskip

The study of the thermodynamic limit for atoms and molecules (i.e. when the minimization is restricted to a specific charge sector) is clearly out of the scope of this paper. Let us however indicate briefly what should be obtained in that case.

We fix some integer $N$ and define the minimization problems in charge sectors:
\begin{equation}
E_L^N(\phi):=\inf\left\{\E_L^\phi(\gamma),\gamma\in\cG_\Lambda^L,\ \tr(\gamma)=N\right\},
\label{min_charge_torus}
\end{equation}
\begin{equation}
E_{\rm BDF}^N(\phi):=\inf\left\{\E_{\rm BDF}^\phi(Q)\ |\ Q\in\mathcal{Q}_\Lambda,\ \tr_{\cP^0_-}(Q)=N\right\}.
\label{min_charge_BDF}
\end{equation}
A minimizer $\tilde Q=\tilde\gamma-\gamma^0$ of \eqref{min_charge_BDF}, if it exists, is a solution of a self-consistent equation of the form \cite{HLS3}
\begin{equation}
\left\{\begin{array}{l}
\displaystyle\tilde\gamma  = -\frac{\sgn(\cD_{\tilde\gamma}-\mu)}{2}\\
\displaystyle{\cD_{\tilde\gamma}}:= D^0+\alpha\left(\rho_{\tilde\gamma}-n\right)\ast\frac{1}{|\cdot|}-\alpha\frac{\tilde\gamma(x,y)}{|x-y|},
\end{array} \right.
\label{scf_eq_molecules}
\end{equation}
$\mu$ being an Euler-Lagrange multiplier due to the charge constraint, interpreted as a chemical potential. Written in terms of $\tilde\cP_-=\tilde\gamma+I_\Lambda/2$, \eqref{scf_eq_molecules} reads
\begin{equation}
\tilde\cP_-=\chi_{(-\ii;0)}\left(\cD_{\tilde\gamma}-\mu\right)=\chi_{(-\ii;\mu)}\left(\cD_{\tilde\gamma}\right).
\end{equation}
We refer the reader to \cite{CI} and \cite[Remark 6]{HLS1} for comments in connection with the Dirac-Fock equations \cite{ES}.

We now \emph{conjecture} that
\begin{equation}
\lim_{L\to\ii}\{E^N_L(\phi)-E_L(0)\}=E^N_{\rm BDF}(\phi)
\label{th_limit_mol}
\end{equation}
which would also justify the Bogoliubov-Dirac-Fock approach for the minimization in charge sectors.

\subsection{The Renormalization Problem}\label{sec_ren}
Throughout this paper, the cut-off $\Lambda$ has been considered as a fixed parameter. However, most of the objects that we have obtained, like for instance the density matrix $\gamma$, are divergent in $\Lambda$. This phenomenon is also encountered in full QED.

To give an example of this, let us consider $g_0(0)$, where $g_0$ has been defined in Theorem \ref{free_vac}. Writing the self-consistent equation \eqref{eq_gamma}, it can be seen that \cite[Eq. $(20)$]{LSie}
$$g_0(0) = m_0+\frac{\alpha}{4\pi^2}\int_{B(0,\Lambda)}\frac{1}{|q|^2}\frac{g_0(|q|)}{\sqrt{g_1(|q|)^2+g_0(|q|)^2}}dq.$$
Therefore, using \eqref{propD3}, one obtains
\begin{eqnarray}
g_0(0)  & \geq & m_0+\frac{\alpha}{4\pi^2}\int_{B(0,\Lambda)}\frac{1}{|q|^2}\frac{m_0g_0(|q|)}{g_0(|q|)\sqrt{|q|^2+m_0^2}}dq\nonumber\\
 & = & m_0\left( 1+\frac{\alpha}{4\pi^2}\int_{B(0,\Lambda)}\frac{1}{|q|^2}\frac{1}{\sqrt{|q|^2+m_0^2}}dq\right)\nonumber\\
 & = & m_0\left(1+\frac{\alpha}{\pi}{\rm arcsinh}(\Lambda/m_0)\right),\label{estim_mass}
\end{eqnarray}
which shows that $g_0(0)$ is divergent as $\Lambda\to\ii$.

In regular QED, the divergences of the (appropriately defined) physical measurable quantities are usually eliminated by means of a \emph{mass} and a \emph{charge renormalization}. The main idea is to assume that the parameters $\alpha$ and $m_0$ appearing in the theory are indeed \emph{bare parameters} which are not physically observable. The \emph{physical parameters} are assumed to be functions of $\alpha$, $m_0$ and the cut-off $\Lambda$
$$\alpha_{\rm ph}=\alpha_{\rm ph}(\alpha,m_0,\Lambda),\qquad  m_{\rm ph}=m_{\rm ph}(\alpha,m_0,\Lambda)$$
and equal the physical values obtained in experiment. These functions should be inverted in order to express the unknown bare quantities in term of the physical quantities
\begin{equation}
\alpha=\alpha(\alpha_{\rm ph},m_{\rm ph},\Lambda),\qquad  m_0=m_0(\alpha_{\rm ph},m_{\rm ph},\Lambda).
\label{def_bares_quantities}
\end{equation}
Using these functions, one expects to
remove (in some sense that needs to be precised) all divergences from physically measurable quantities.

 Mass and charge renormalization however does not
remove all divergences in the theory. Certain quantities, e.g. the
bare Feynman propagator $S_F$ (either at equal times or at general space
time points), are still divergent. The expectation is that all these
divergences cancel in physically measurable quantities and that they are
therefore of no real relevance in formulating the theory.

Although there is no real need to do this, it is often convenient to
introduce a renormalization of the bare Feynman propagator $S_F$. This is
referred to as a \emph{wavefunction renormalization}. In full QED \cite{Dy2} it is claimed
that the divergence in the Feynman propagator may be removed by a
\emph{multiplicative renormalization} and that the renormalized propagator has
the same pole near mass shell in 4-momentum space as a free propagator
corresponding to a particle with the correct physical mass.

Note that in practice, this theoretical renormalization procedure is always used to justifying the dropping of the divergent terms obtained at each order of the perturbation theory \cite{Dy2}. For this fact to be true, it is particularly important that renormalization can be expressed by means of \emph{multiplicative} parameters in front of the different propagators \cite{Dy2}.

In Hartree-Fock QED, it is not clear at all if the usual renormalization program of QED can be applied (especially when photons are not included). In \cite{HLS2}, a renormalization of the charge is proposed but it seems to be only valid if the exchange term is neglected. In  \cite[p. 194--195]{RGA}, it is argued that mass and charge renormalization is alone not enough to completely remove the divergences of the HF theory by means of multiplicative parameters.

In \cite{LSie}, it is proposed, among other possibilities, to write the operator of Theorem \ref{free_vac} as
$$\cD^0(p)=\frac{g_1(|p|)}{|p|}\left(\alp\cdot p+\frac{|p|g_0(|p|)}{g_1(|p|)}\beta\right)$$
and to interpret $g_0(0)/g_1'(0)$ as being the physical mass $m_{\rm ph}$ and $g_1'(0)$ as being the wavefunction renormalization constant. This proposal does not at all correspond to the procedure described above.
Indeed, multiplying $\cD^0$ by a constant does not change the equal time Feynman propagator $\gamma^0$, since $\gamma^0=-\sgn(\cD^0)/2$.

We believe that the correct solution is rather to define the quantities $\alpha_{\rm ph}$ and $m_{\rm ph}$ for a free Hartree-Fock electron (strictly speaking, photons should be included). Namely, we define the physical mass as the minimum energy of a free electron which means, using \eqref{min_charge_BDF},
\begin{equation}
m_{\rm ph}:=E_{\rm BDF}^1(0).
\label{m_ph}
\end{equation}
Since an electron never sees its own field but interacts with the Dirac sea, we believe that $E_{\rm BDF}^1(0)$ possesses a minimizer $\tilde Q=\tilde\gamma-\gamma^0$, i.e. that an electron can bind alone when interacting with the Dirac sea (this has been first proposed by \'Eric Séré). According to \eqref{scf_eq_molecules}, the operator $\tilde\gamma$ would then be a solution of the self-consistent equation
\begin{equation}
\left\{\begin{array}{l}
\displaystyle\tilde\gamma  = -\frac{\sgn(\cD_{\tilde\gamma}-\mu)}{2}\\
\displaystyle{\cD_{\tilde\gamma}}:= D^0+\alpha\rho_{\tilde\gamma}\ast\frac{1}{|\cdot|}-\alpha\frac{\tilde\gamma(x,y)}{|x-y|},
\end{array} \right.
\label{bind_electron}
\end{equation}
where $\mu$ is a chemical potential chosen such that $\tr_{\cP^0_-}(\tilde\gamma-\gamma^0)=1$.
Formally, we could then define $\alpha_{\rm ph}$ by
\begin{equation}
\alpha_{\rm ph}:=\alpha \lim_{x\to\ii}|x|\left(\rho_{\tilde\gamma}\ast{|\cdot|^{-1}}\right)(x).
\label{alpha_ph}
\end{equation}
If $\widehat{\rho_{\tilde\gamma}}$ is smooth enough, for instance $\widehat{\rho_{\tilde\gamma}}\in W^{1,\ii}(B(0,\Lambda))$, this is equivalent to
$$\alpha_{\rm ph}:=(2\pi)^{-3/2}\alpha\widehat{\rho_{\tilde\gamma}}(0).$$
If $\tilde\gamma-\gamma^0$ were trace-class, one would have $(2\pi)^{-3/2}\widehat{\rho_{\tilde\gamma}}(0)=\tr(\tilde\gamma-\gamma^0)=\int_{\R^3}\rho_{\tilde\gamma-\gamma^0}=1$ and $\alpha_{\rm ph}=\alpha$. But in the present case, it is known that when $\tilde\gamma$ is a solution of the self-consistent equation \eqref{bind_electron} such that $\rho_{\tilde\gamma}\neq0$, then $\tilde\gamma-\gamma^0$ is \emph{never trace-class} \cite{KS,HS,HLS1}. Expanding \eqref{bind_electron} to first order in the bare $\alpha$, this should rather lead to a formula of the form \cite{HLS2}
$$\alpha_{\rm ph}(\alpha,m_0,\Lambda)=\frac{\alpha}{1+\frac{2\alpha}{3\pi}\log(\Lambda/m_0)}+o(\alpha),$$
the usual formula in QED \cite[Eq. $(7.18)$]{IZ}.

Both \eqref{m_ph} and \eqref{alpha_ph} would define $m_{\rm ph}$ and $\alpha_{\rm ph}$ as extremely complicated non-linear functions of $\alpha$, $m_0$ and $\Lambda$. A challenging task is to study the finiteness of measurable quantities like for instance the energy of an electron in the presence of an  external field $E_{\rm BDF}^1(\phi)$, when $\alpha_{\rm ph}$ and $m_{\rm ph}$ are fixed to be the observed physical quantities.

\section{Proofs}\label{proofs}
\subsection{Preliminaries}
We shall need the following lemma, which summarizes the properties of $W_L$, defined in \eqref{def_Coulomb}:
\begin{lemma}\label{prop_Coulomb} The potential $W_L$ satisfies the following properties:
\begin{enumerate}
\item there exists a constant $C$ such that \cite{LS}
\begin{equation}
\norm{W_L(x)-\frac{1}{|x|}}_{L^\ii(\cC_L)}\leq \frac{C}{L} ;
\label{conv_Coulomb}
\end{equation}
\item (Kato-type inequality) for any $m_0>0$ and $L$ large enough, there exists a constant $C_\Lambda^L(m_0)$ such that, on $\gH_\Lambda^L$,
\begin{equation}
W_L\leq C_\Lambda^L(m_0)|D^0|,\qquad \lim_{L\to\ii}C_\Lambda^L(m_0)=C_\Lambda(m_0)
\label{Kato}
\end{equation}
where $C_\Lambda(m_0)\leq \pi/2$ is the optimal constant for the inequality
$$\frac{1}{|x|}\leq C_\Lambda(m_0) |D^0|$$
on $\gH_\Lambda$.
\end{enumerate}
\end{lemma}
\begin{proof}
We know from \cite[Eq. $(112)$]{LS} that $C:=\norm{W_1-\frac{1}{|\cdot|}}_{L^\ii(\cC_1)}<\ii$. On the other hand, we have the scaling property
$$W_L(x)=\frac{1}{L}W_1\left(\frac{x}{L}\right)$$
which implies that
$$\norm{W_L-\frac{1}{|\cdot|}}_{L^\ii(\cC_L)}\leq \frac{C}{L}.$$

For a fixed bare mass $m_0>0$, let us define (we do not mention $m_0$ for simplicity)
$$C_\Lambda^L:=\sup_{\substack{\psi\in \gH_\Lambda^L\\ \pscal{|D^0|\psi,\psi}\leq 1}}\int_{\cC_L}W_L(x)|\psi(x)|^2dx<\ii,$$
$$C_\Lambda:=\sup_{\substack{\psi\in \gH_\Lambda\\ \pscal{|D^0|\psi,\psi}\leq 1}}\int_{\R^3}\frac{|\psi(x)|^2}{|x|}dx<\ii.$$
A proof that $C_\Lambda^L$ is finite can be found in \cite{CP}.
Let us consider a function $\psi\in\gH_\Lambda$ such that $\widehat{\psi}$ is continuous on the ball $B(0,\Lambda)$ and
$$\left|\int_{\R^3}\frac{|\psi(x)|^2}{|x|}dx-C_\Lambda\right|\leq \epsilon,\qquad \pscal{|D^0|\psi,\psi}<1.$$
We now approximate $\psi$ by a sequence of functions in $\gH_\Lambda^L$ by defining
$$\psi_L(x)=\frac{(2\pi)^{3/2}}{L^3}\sum_{k\in\Gamma_\Lambda^L}\widehat{\psi}(k)e^{ik\cdot x}.$$
We have
$$\norm{\psi_L}_{L^\ii(\cC_L)}\leq \frac{(2\pi)^{3/2}|\Gamma_\Lambda^L|}{L^3}\norm{\widehat\psi}_{L^\ii(B(0,\Lambda))},$$
$$\norm{\nabla\psi_L}_{L^\ii(\cC_L)}\leq \frac{\Lambda(2\pi)^{3/2}|\Gamma_\Lambda^L|}{L^3}\norm{\widehat\psi}_{L^\ii(B(0,\Lambda))}.$$
Therefore $(\psi_L)_L$ is bounded in $W^{1,\ii}(\R^3)$ and $\psi_L\to\psi$ uniformly on compact subsets of $\R^3$. On the other hand,
$$\norm{\psi_L}_{L^2(\cC_L)}^2\leq \frac{(2\pi)^3|\Gamma_\Lambda^L|}{L^3}\norm{\widehat\psi}_{L^\ii(B(0,\Lambda))}^2$$
which is also uniformly bounded in $L$.
Using now
$$\left|\int_{\cC_L}|\psi_L(x)|^2W_L(x)\,dx-\int_{\cC_L}\frac{|\psi_L(x)|^2}{|x|}dx\right|\leq \norm{\psi_L}_{L^2(\cC_L)}^2\norm{W_L-\frac{1}{|\cdot|}}_{L^\ii(\cC_L)}\leq \frac{C}{L},$$
we easily deduce that
$$\lim_{L\to\ii}\int_{\cC_L}|\psi_L(x)|^2W_L(x)dx=\int_{\R^3}\frac{|\psi(x)|^2}{|x|}dx.$$
On the other hand,
$$\pscal{|D^0|\psi_L,\psi_L}=\left(\frac{2\pi}{L}\right)^3\sum_{k\in\Gamma_\Lambda^L}\pscal{|D^0(k)|\widehat\psi(k),\widehat\psi(k)}_{\C^4}$$
which converges to $\pscal{|D^0|\psi,\psi}$ as $L\to\ii$, and therefore $\pscal{|D^0|\psi_L,\psi_L}\leq 1$ for $L$ large enough. This shows that
$$\liminf_{L\to\ii}C_\Lambda^L\geq C_\Lambda+\epsilon$$
for any $\epsilon>0$ and thus $\liminf_{L\to\ii}C_\Lambda^L\geq C_\Lambda$.

Since $\gH_\Lambda^L$ is finite dimensional, there exists for any $L$ an optimal function $f_L\in\gH_\Lambda^L$ satisfying $\pscal{|D^0|f_L,f_L}\leq 1$ and $\int_{\cC_L}W_L(x)|f_L(x)|^2dx=C_\Lambda^L$. Thanks to the cut-off in Fourier space, $(f_L)_L$ is bounded for instance in $H^1(\cC_L)$ and therefore $f_L\to f\in\gH_\Lambda$ in $L^2_{\rm loc}(\R^3)$, up to a subsequence. Passing to the limit like above, we obtain
$$\lim_{L\to\ii}\int_{\cC_L}|f_L(x)|^2W_L(x)\,dx=\int_{\R^3}\frac{|f(x)|^2}{|x|}\,dx$$
and $\pscal{|D^0|f,f}\leq 1$. This shows that
$$\lim_{L\to\ii}C_\Lambda^L\leq \int_{\R^3}\frac{|f(x)|^2}{|x|}\,dx\leq C_\Lambda.$$
As a conclusion, we have
$$\lim_{L\to\ii}C_\Lambda^L=C_\Lambda\leq \pi/2$$
which ends the proof of Lemma \ref{prop_Coulomb}.
\end{proof}

\subsection{Proof of Theorem \ref{free_vac}}\label{proof_free_vac}
The functional $\T$ can as well be written
$$\T(f)=\frac{1}{(2\pi)^{3}}\left(\int_{B(0,\Lambda)}\tr_{\C^4}[D^0(p)f(p)]dp-\frac{\alpha}{2}\int_{\R^3}\frac{|\check{f}(x)|^2}{|x|}dx\right),$$
which easily shows that $\T$ is continuous for the weak topology of $L^2(B(0,\Lambda),\S_4(\C))$. Since $\A_\Lambda$ is bounded and closed in this topology, $\T$ possesses a minimizer on $\A_\Lambda$. In order to show the uniqueness and the mentioned properties of this minimizer, we shall now minimize $\T$ on a special subset of $\A_\Lambda$.

\medskip

\noindent {\bf $\bullet$ Step 1: }{\it Minimization of $\T$ on a restricted set.} Let us introduce the following subset of $\A_\Lambda$
\begin{equation}
\B_\Lambda:=\left\{f\in\A_\Lambda\ |\ \exists (f_0,f_1)\in G,\ f(p)=\alp\cdot\omega_p f_1(|p|)+f_0(|p|)\beta\right\}
\label{B_Lambda}
\end{equation}
where $\omega_p=p/|p|$ and $G\subset L^\ii([0;\Lambda],\R)^2$ is the set of pairs $(f_0,f_1)$ satisfying the following properties
\begin{eqnarray}
 & f_0\leq0 \text{ and } f_1\leq 0,\label{prop1}\\
 & f_0^2+f_1^2\leq 1/4\label{prop2}\\
 & \forall p\in B(0,\Lambda),\quad \displaystyle |p|\int_{B(0,\Lambda)}\frac{f_0(|q|)}{|p-q|^2}dq\leq m_0\int_{B(0,\Lambda)}\frac{\omega_p\cdot\omega_qf_1(|q|)}{|p-q|^2}dq.\label{prop3}
\end{eqnarray}
Notice that $G$ is a bounded convex set, closed in the weak topology of $L^2([0;\Lambda],\R)^2$. Therefore, $\B_\Lambda$ is also a closed subset of $\A_\Lambda$. As a consequence, we deduce that $\T$ possesses a minimizer $\bar f$ on $\B_\Lambda$. Let us now show that $\bar f$ satisfies the same Euler-Lagrange equation as global minimizers of $\T$ on $\A_\Lambda$.

Indeed, considering perturbations of the form $(1-t)\bar f+t g$ for any $g\in\B_\Lambda$, $t\in[0;1]$, one easily shows that $\bar f $ is a minimizer of the following energy on $\B_\Lambda$
\begin{equation}
\T_{\bar f}(g)=\int_{B(0,\Lambda)}\tr_{\C^4}\left(\cD^0(p)g(p)\right)dp,
\label{def_energy_perturb}
\end{equation}
i.e. $\T_{\bar f}(g)\geq \T_{\bar f}(\bar f)$ for any $g\in\B_\Lambda$. Here $\cD^0$ is the mean-field operator associated with $\bar f$
\begin{eqnarray}
\cD^0(p) & = & D^0(p)-\frac{\alpha}{2\pi^2}\int_{B(0,\Lambda)}\frac{\bar f(q)}{|p-q|^2}dq\label{def_new_D0}\\
 & = & \alp\cdot\omega_p\left(|p|-\frac{\alpha}{2\pi^2}\int_{B(0,\Lambda)}\frac{\omega_p\cdot\omega_q\bar f_1(|q|)}{|p-q|^2}dq\right)\nonumber\\
 & & \qquad\qquad + \beta\left(m_0-\frac{\alpha}{2\pi^2}\int_{B(0,\Lambda)}\frac{\bar f_0(|q|)}{|p-q|^2}dq\right).\nonumber\\
 & = & \alp\cdot\omega_pg_1(|p|)+\beta g_0(|p|),\nonumber
\end{eqnarray}
with
\begin{eqnarray}
g_1(|p|) & = & |p|-\frac{\alpha}{2\pi^2}\int_{B(0,\Lambda)}\frac{\omega_p\cdot\omega_q\bar f_1(|q|)}{|p-q|^2}dq\nonumber \\
 & = & |p|-\frac{\alpha}{\pi}\int_0^\Lambda\frac{v}{|p|}Q_1\left(\frac12\left(\frac{|p|}{v}+\frac{v}{|p|}\right)\right)\bar f_1(v)dv,\label{def_g1}
\end{eqnarray}
and
\begin{eqnarray}
g_0(|p|) & = & m_0-\frac{\alpha}{2\pi^2}\int_{B(0,\Lambda)}\frac{\bar f_0(|q|)}{|p-q|^2}dq\nonumber\\
 & = & m_0-\frac{\alpha}{\pi}\int_0^\Lambda\frac{v}{|p|}Q_0\left(\frac12\left(\frac{|p|}{v}+\frac{v}{|p|}\right)\right)\bar f_0(v)dv.\label{def_g0}
\end{eqnarray}
Here $Q_0$ and $Q_1$ are positive functions on $(1;\ii)$, defined in \cite[Eq. $(24)$ and $(25)$]{LSie}, arising from the integration of the angle as shown in \cite{LSie}.

Due to the assumptions of \eqref{prop1}, we deduce that $g_1(|p|)\geq|p|$ and $g_0(|p|)\geq m_0$, which implies
$$|\cD^0(p)|\geq \sqrt{|p|^2+m_0^2}\geq m_0>0.$$
Therefore $\cD^0(p)$ is always invertible. This now easily implies that the unique \emph{global} minimizer of $\T_{\bar f}$ defined in \eqref{def_energy_perturb}, on $\A_\Lambda$, is
$$\bar f'(p)=-\frac{{\rm sgn}(\cD^0(p))}{2}=-\frac{\cD^0(p)}{2|\cD^0(p)|}.$$
We now show that indeed $\bar f'\in\B_\Lambda$, which will imply $\bar f' =\bar f$. To this end, we just have to prove that
$$\bar f'_0(p)=-\frac{g_0(p)}{2|\cD^0(p)|}\quad \text{ and }\quad \bar f'_1(p)=-\frac{g_1(p)}{2|\cD^0(p)|}$$
satisfy the additional properties \eqref{prop1} and \eqref{prop3}. For \eqref{prop1}, this is an immediate consequence of the fact that $Q_0$ and $Q_1$ are non-negative on $(1;\ii)$, whereas \eqref{prop3} is proved in \cite[Theorem 2]{LSie}.

As a conclusion, we have found a minimizer of $\T$ on the restricted set $\B_\Lambda$ which is a solution to the self-consistent equation
$$\bar f(p)=-\frac{{\rm sgn}(\cD^0(p))}{2}$$
where $\cD^0(p)$ is defined in \eqref{def_new_D0}. Notice that by construction, $\cD^0(p)$ satisfies the properties \eqref{propD1}, \eqref{propD2} and \eqref{propD3}. Of course, $\gamma^0:=\bar f(-i\nabla)$ and $\cP^0_-:=\gamma^0+I_\Lambda/2$ are respectively solutions of the self-consistent equations \eqref{scf_proj} and \eqref{scf_proj2}.

\medskip

\noindent {\bf $\bullet$ Step 2: }{\it the minimizer $\bar f$ of $\T$ on $\B_\Lambda$ is its unique global minimizer on $\A_\Lambda$.} We now show that we not only have solved the self-consistent equation \eqref{scf_proj}, but that $\bar f$ is the unique global minimizer of $\T$.

To this end, we compute, for some $f\in\A_\Lambda$,
$$\T(f)=\T(\bar f)+\frac{1}{(2\pi)^{3}}\left(\int_{B(0,\Lambda)}\tr_{\C^4}[\cD^0(p)Q(p)]dp-\frac{\alpha}{2}\int_{\R^3}\frac{|\check Q(x)|^2}{|x|}\,dx\right),$$
where $Q(p)=f(p)-\bar f(p)$. Since $-I_{\C^4}/2\leq f(p)\leq I_{\C^4}/2$ due to \eqref{prop2} and $\bar f(p)=(\cP^0_-(p)-\cP^0_+(p))/2$, we see that
\begin{equation}
\forall p\in B(0,\Lambda),\qquad -\cP^0_-(p)\leq Q(p)\leq \cP^0_+(p).
\label{contrainte_Q}
\end{equation}
We now adapt arguments from \cite{BBHS,HLS1} in the translation-invariant case. Namely, \eqref{contrainte_Q} implies
$$Q(p)^2\leq \cP^0_+(p)Q(p)\cP^0_+(p)-\cP^0_-(p)Q(p)\cP^0_-(p)$$
and
$$0\leq\tr_{\C^4}(|\cD^0(p)|Q(p)^2)\leq \tr_{\C^4}(\cD^0(p)Q(p))$$
for any $p\in B(0,\Lambda)$.
Now, by Kato's inequality,
$$\int_{\R^3}\frac{|\check Q(x)|^2}{|x|}\,dx\leq \frac\pi2\int\tr_{\C^4}(|p|\,Q(p)^2)dp\leq \frac\pi2\int\tr_{\C^4}(|\cD^0(p)|\,Q(p)^2)dp$$
for
$$|p|\leq \sqrt{|p|^2+m_0^2}\leq |\cD^0(p)|$$
by \eqref{propD3}, and therefore, when $0\leq \alpha<4/\pi$,
\begin{eqnarray*}
\T(f) & \geq & \T(\bar f)+\frac{1}{(2\pi)^{3}}(1-\alpha\pi/4)\int_{B(0,\Lambda)}\tr_{\C^4}[|\cD^0(p)|(f-\bar f)^2(p)]dp\\
& \geq & \T(\bar f)+\frac{(1-\alpha\pi/4)m_0}{(2\pi)^{3}}\norm{f-\bar f}_{L^2(B(0,\Lambda),\S_4(\C))}^2.
\end{eqnarray*}
This immediately implies that $\bar f$ is the unique global minimizer of $\T$ and also shows that any minimizing sequence $(f_n)\subset\A_\Lambda$ necessarily satisfies  $f_n\to\bar f$ strongly in $L^2(B(0,\Lambda),\S_4(\C))$.

\medskip

\noindent {\bf $\bullet$ Step 3: }{\it regularity of $\cD^0(p)$.} It remains to show that the self-consistent free Dirac operator $\cD^0$ is smooth in the Fourier domain. To this end, we notice that it satisfies the equation
$$\cD^0(p)=D^0(p)-\frac{\alpha}{4\pi^2}\int_{B(0,\Lambda)}\frac{\cD^0(q)}{|\cD^0(q)|\;|p-q|^2}dq$$
or, on $B(0,\Lambda)$,
$$\cD^0=D^0-\frac{\alpha}{4\pi^2}\frac{\cD^0}{|\cD^0|}\ast\frac{1}{|\cdot|^2}.$$
Therefore
$$|\nabla|\cD^0=|\nabla| D^0-\frac{\alpha}{2}\frac{\cD^0}{|\cD^0|}$$
which shows that $\cD^0\in H^{1}(B(0,\Lambda))$. Applying now $\nabla$ to this equation and using a easy boot-strap argument, one obtains that
$$\cD^0\in \bigcap_{m\geq1}H^m(B(0,\Lambda))\subset\mathcal{C}^\ii(\overline{B(0,\Lambda)}).$$
which ends the proof of Theorem \ref{free_vac}.\qed

\subsection{Proof of Theorem \ref{thermo1}}\label{proof_thermo1}
Our proof proceeds as follows: we first minimize $\E^0_L$ on a restricted set of translation invariant operators and study the limit of the corresponding sequence of minimizers as $L\to\ii$. Using then the properties of the solution $\bar f$ constructed in Theorem \ref{free_vac}, we show that, for each $L$ large enough, the so-obtained local minimizer is indeed the unique global minimizer of $\E_L^0$.

\medskip

\noindent {\bf $\bullet$ Step 1: }{\it Definition of the restricted minimization procedure.} Let us define the following minimization problem\footnote{Recall that we denote also by $-i\nabla$ the operator acting on $\gH_\Lambda^L$, i.e. the multiplication operator in the Fourier domain by $(k)_{k\in\Gamma_\Lambda^L}$. Therefore $f(-i\nabla)$ is just the multiplication operator by $(f(k))_{k\in\Gamma_\Lambda^L}$, in the Fourier domain.}:
\begin{equation}
E_L^\T:=\inf\left\{\E_L^0(\gamma)\ |\ \gamma=f(-i\nabla),\ f\in\B_\Lambda^L\right\}
\label{min_tr_box}
\end{equation}
where $\B_\Lambda^L$ is the set of functions $f:\Gamma_\Lambda^L\mapsto \S_4(\C)$ which take the special form (compare with $\B_\Lambda$ defined in \eqref{B_Lambda})
\begin{equation}
\forall p\in\Gamma_\Lambda^L,\qquad f(p)=\alp\cdot A(p)+\beta b(p)
\label{form_f_1}
\end{equation}
where $A:\Gamma_\Lambda^L\mapsto \R^3$ and $b:\Gamma_\Lambda^L\mapsto\R$ are such that
\begin{equation}
\forall p\in\Gamma_\Lambda^L,\quad  b(p)\leq 0,
\end{equation}
\begin{equation}
\forall p\in\Gamma_\Lambda^L,\quad  |A(p)|^2+b(p)^2\leq {1}/{4}.
\end{equation}
The energy of such a state reads
\begin{eqnarray}
\quad\quad\quad \E_L^0(\gamma) & = & \tr(D^0\gamma)-\frac\alpha2\iint_{(\bT_L)^2}|\gamma(x,y)|^2W_L(x-y)dx\,dy\label{energy_rest_min}\\
 & = & \sum_{p\in\Gamma_\Lambda^L}\tr_{\C^4}(D^0(p)f(p)) - \frac{\alpha}{2(2\pi)^{3}}L^3\int_{\bT_L}|\check{f}(x)|^2W_L(x)dx\nonumber\\
 & = & \!\sum_{p\in\Gamma_\Lambda^L}\tr_{\C^4}(D^0(p)f(p)) - \frac{\alpha(2\pi)^{3/2}}{2L^3}\!\!\sum_{p,q\in\Gamma_\Lambda^L}\tr_{\C^4}(f(p)f(q))\widehat{W_L}(p-q),\nonumber
\end{eqnarray}
where we have used that, due to the special form \eqref{form_f_1} of $f$, the density of $\gamma$ vanishes,
$\rho_\gamma=L^{-3}\sum_{p\in\Gamma_\Lambda^L}\tr_{\C^4}f(p)=0$.

\medskip

\noindent {\bf $\bullet$ Step 2: }{\it Upper bound on the energy per unit volume.} Let us now show that
$$\limsup_{L\to\ii}\frac{E^\T_L}{L^3}\leq E^\T=\min\left\{\T(f)\ |\ f\in\A_\Lambda\right\}.$$
This is easily done by considering the minimizer $\bar f$ of $\T$ as a test function. For the sake of clarity, we denote by $g_L$ the restriction of $\bar f$ to $\Gamma_\Lambda^L$ and by $G_L=g_L(-i\nabla)$ the associated operator acting on $\gH_\Lambda^L$, which is the multiplication operator in the Fourier domain by $(\bar f(k))_{k\in\Gamma_\Lambda^L}$. Notice that
\begin{equation}
\check{g_L}(x)=\frac{(2\pi)^{3/2}}{L^3}\sum_{k\in\Gamma_\Lambda^L}\bar f(k)e^{ik\cdot x}
\label{hat_g_L}
\end{equation}
and that $g_L=\bar f_{|\Gamma_\Lambda^L}$ belongs to $\B_\Lambda^L$ for any $L$ due to the properties of $\bar f$ stated in Theorem \ref{free_vac}.

Using $|\bar f(k)|\leq 1/2$ for any $k\in B(0,\Lambda)$ and \eqref{hat_g_L}, one sees that
\begin{equation}
\norm{\check{g_L}}_{L^\ii(\R^3)}\leq \frac{(2\pi)^{3/2}|\Gamma_\Lambda^L|}{2L^3},\qquad
\norm{\nabla\check{g_L}}_{L^\ii(\R^3)}\leq \frac{\Lambda(2\pi)^{3/2}|\Gamma_\Lambda^L|}{2L^3}.
\label{estim_norm1}
\end{equation}
Therefore, $(\check{g_L})_L$ is bounded in $W^{1,\ii}(\R^3)$ and converges to $\check{\bar f}$ uniformly on compact subsets of $\R^3$. Finally, we have $\rho_{G_L}\equiv0$ since $\tr_{\C^4}\bar f(k)=0$ for any $k\in B(0,\Lambda)$ and therefore, by \eqref{energy_rest_min},
$$\frac{\E_L^0(G_L)}{L^3}= \frac{1}{L^3}\sum_{k\in\Gamma_\Lambda^L}\tr_{\C^4}(D^0(k)\bar f(k))-\frac{\alpha}{2(2\pi)^3}\int_{\bT_L}|\check{g_L}(x)|^2W_L(x)dx.$$
Arguing like in the proof of Lemma \ref{prop_Coulomb}, one easily shows that
$$\lim_{L\to\ii}\frac{\E_L^0(G_L)}{L^3}=\T(\bar f),$$
hence
$$\limsup_{L\to\ii}\frac{E^\T_L}{L^3}\leq \T(\bar f)=E^\T.$$

\medskip

\noindent {\bf $\bullet$ Step 3: }{\it Existence of a minimizer for $E_L^\T$.} In order to show the converse inequality
\begin{equation}
\liminf_{L\to\ii}\frac{E^\T_L}{L^3}\geq E^\T,\label{lim_inf_rest_energy}
\end{equation}
we first prove the existence of a minimizer for \eqref{min_tr_box}.

Since $\gH_\Lambda^L$ is finite-dimensional, there exists a solution $\gamma^0_L=f_L^0(-i\nabla)$, $f^0_L\in\mathcal{B}_\Lambda^L$, of the minimization problem \eqref{min_tr_box}. We now argue like in the proof of Theorem \ref{free_vac} (step 1) to show that $\gamma_L^0$ satisfies the same equation as global minimizers of $\E_L^0$ on $\cG_\Lambda^L$. Due to the fact that $\B_\Lambda^L$ is convex, $f_L^0$ also minimizes
\begin{equation}
\T_{f_L^0}(f)= \sum_{p\in\Gamma_\Lambda^L}\tr_{\C^4}(\cD^0_L(p)f(p))
\label{T_f_L}
\end{equation}
in $\B_\Lambda^L$, where $\cD^0_L$ is the mean-field operator associated with $f_L^0=\alp\cdot A_L^0+\beta b_L^0$:
\begin{eqnarray*}
\cD^0_L(p) & = & D^0(p)-\frac{\alpha(2\pi)^{3/2}}{L^3}\sum_{q\in\Gamma_\Lambda^L}f_L^0(q)\widehat{W_L}(q-p)\\
 & = & \alp\cdot\left(p-\frac{\alpha(2\pi)^{3/2}}{L^3}\sum_{q\in\Gamma_\Lambda^L}A_L^0(q)\widehat{W_L}(q-p)\right)\\
 & & \qquad +\beta \left(m_0-\frac{\alpha(2\pi)^{3/2}}{L^3}\sum_{q\in\Gamma_\Lambda^L}b^0_L(q)\widehat{W_L}(q-p)\right).
\end{eqnarray*}
Notice that
\begin{equation}
m_0-\frac{\alpha(2\pi)^{3/2}}{L^3}\sum_{q\in\Gamma_\Lambda^L}b^0_L(q)\widehat{W_L}(p-q)\geq m_0
\label{estim_mass_box}
\end{equation}
which shows that $\cD^0_L(p)$ is invertible for all $p$. The unique global minimizer of $\T_{f_L}$ on $\B_\Lambda^L$ is $-{\rm sgn}(\cD^0_L)/2$, which is easily seen to belong to $\B_\Lambda^L$. We therefore obtain, for $k\in\Gamma_\Lambda^L$,
\begin{equation}
f_L^0(k)=-\frac{\cD^0_L(k)}{2|\cD^0_L(k)|}.
\label{eq_f_L}
\end{equation}
Saying differently, $\gamma_L^0:=f_L^0(-i\nabla)$ satisfies the equation
$$\gamma_L^0=-\frac{\sgn(\cD^0_L)}{2}, \qquad \cD^0_L:=D^0-\alpha\gamma_L^0(x,y)W_L(x-y)$$
or, written in terms of the projector  $\cP^0_L:=\gamma_L^0+I_\Lambda^L/2$,
$$\cP_L^0=\chi_{(-\ii;0)}(\cD^0_L).$$

\medskip

\noindent {\bf $\bullet$ Step 4: }{\it Lower bound on the energy per unit volume.} We now prove \eqref{lim_inf_rest_energy} by studying the weak limit of $\gamma^0_L=f_L^0(-i\nabla)$ defined in the previous step.
Using the same type of estimate as \eqref{estim_norm1}, we obtain that $(\check{f_L^0})$ is bounded in $W^{1,\ii}(\R^3)$ and it therefore converges to some function $\check{f}$, uniformly on compact subsets of $\R^3$, up to a subsequence. On the other hand, let us define the stepwise function $\nu_L[f_L^0]$ by
\begin{equation}
{\nu_L[f_L^0]}(p)=\sum_{k\in\Gamma_\Lambda^L}f_L^0(k)\chi_{k}(p),
\label{step}
\end{equation}
$\chi_k$ being the characteristic function of $\prod_{i=1}^3[k_i;k_i+2\pi/L[\cap B(0,\Lambda)$. Then $\nu_L[f_L^0]$ is bounded in $L^\ii(B(0,\Lambda),\S_4(\C))$ and thus $\nu_L[f_L^0]\wto h\in\A_\Lambda$, up to a subsequence, for instance for the weak topology of $L^2(B(0,\Lambda),\S_4(\C))$.
To see that $h=f$, we can for instance notice that
$$\check{f_L^0}(x)-\check{\overbrace{\nu_L[f_L]}}(x)=(2\pi)^{-3/2}\sum_{k\in\Gamma_\Lambda^L}f_L^0(k)\int\chi_k(p)\left(e^{ik\cdot x}-e^{ip\cdot x}\right)dp+O\left(\frac1{L}\right),$$
where our $O(1/L)$ is due to the $k\in\Gamma_\Lambda^L$ which are close to the boundary of $B(0,\Lambda)$. Hence, for any compact subset $K$ of $\R^3$,
\begin{eqnarray*}
\norm{\check{f_L^0}-\check{\overbrace{\nu_L[f_L^0]}}}_{L^\ii(K)} & \leq & C(2\pi)^{-3/2}\sum_{k\in\Gamma_\Lambda^L}|{f_L^0}(k)|\int\chi_k(p)|(k-p)\cdot x|dp+O\left(\frac1{L}\right)\\
 & \leq & \frac{C_K}{L}|B(0,\Lambda)|+O\left(\frac1{L}\right)
\end{eqnarray*}
for some constants $C$ and $C_K$, which shows that $\check{\overbrace{\nu_L[f_L^0]}}\to \check{f}$ on compact subsets of $\R^3$ and implies $f=h\in\A_\Lambda$.

The energy of $\gamma_L^0$ reads
\begin{multline}
\frac{\E_L^0(\gamma_L^0)}{L^3} = \frac{1}{L^3}\sum_{k\in\Gamma_\Lambda^L}\tr_{\C^4}(D^0(k)f_L^0(k))-\frac{\alpha}{2(2\pi)^3}\int_{\bT_L}|\check{f_L^0}(x)|^2W_L(x)dx\\
 +\frac{\alpha}{2}\mu L^2(\rho_{\gamma_L})^2.
\label{energy_min_box}
\end{multline}
We have
$$\lim_{L\to\ii}\int_{\bT_L}|\check{f_L^0}(x)|^2W_L(x)dx=\int_{\R^3}\frac{|\check{f}(x)|^2}{|x|}dx$$
due to the fact that $(\check{f_L^0})_L$ converges locally to $f$ and that
$$\norm{f_L^0}_{L^2(\bT_L)}=\left(\frac{2\pi}{L}\right)^3\sum_{k\in\Gamma_\Lambda^L}|{f_L^0}(k)|^2\leq \frac{(2\pi)^3|\Gamma_\Lambda^L|}{4L^3}$$
is uniformly bounded in $L$. On the other hand, using the notation \eqref{step},
$$\frac{1}{L^3}\sum_{k\in\Gamma_\Lambda^L}\tr_{\C^4}(D^0(k)f_L^0(k))=(2\pi)^{-3}\int_{B(0,\Lambda)}\tr_{\C^4}(\nu_L[D^0](p)\nu_L[f_L^0](p))dp+O\left(\frac1{L}\right).$$
Since $\nu_L[f_L^0]\wto f$ weakly and $\nu_L[D^0]\to D^0$ strongly in $L^2(B(0,\Lambda))$, we deduce that
$$\lim_{L\to\ii}\frac{1}{L^3}\sum_{k\in\Gamma_\Lambda^L}\tr_{\C^4}(D^0(k)f_L^0(k))=(2\pi)^{-3}\int_{B(0,\Lambda)}\tr_{\C^4}(D^0(p)f(p))dp$$
and
$$\lim_{L\to\ii}\left(\frac{1}{L^3}\sum_{k\in\Gamma_\Lambda^L}\tr_{\C^4}(D^0(k)f_L^0(k))-\frac{\alpha}{2(2\pi)^3}\int_{\bT_L}|\check{f_L^0}(x)|^2W_L(x)dx\right)=\T(f).$$
Recall that $\mu>0$, which implies due to \eqref{energy_min_box}
$$\liminf_{L\to\ii}\frac{E^\T_L}{L^3}\geq \T(f)\geq E^\T$$
where we have used that $f\in\A_\Lambda$. This of course shows that
$$\lim_{L\to\ii}\frac{E^\T_L}{L^3}=E^\T.$$
and, as a matter of fact, $\T(f)=E^\T$. By Theorem \ref{free_vac}, $\bar f$ is the unique minimizer of $\T$ on $\A_\Lambda$, and thus $f=\bar f$.

\medskip

\noindent {\bf $\bullet$ Step 5: }{\it Study of the convergence of $\gamma_L^0=f_L^0(-i\nabla)$.}
We know that  $\nu_L[f_L^0]\wto \bar f$ weakly in $L^2(B(0,\Lambda))$ and that $\check{f_L^0}\to\check{\bar f}$ uniformly on compact subsets of $\R^3$. Notice that $\nu_L[f_L^0]\in\A_\Lambda$ for any $L$. Since we know from the proof of Theorem \ref{free_vac} that $\T$ is continuous on $\A_\Lambda$ for the weak topology of $L^2(B(0,\Lambda))$, we obtain
$$\lim_{L\to\ii}\T(\nu_L[f_L^0])=\T(\bar f),$$
which means that $(\nu_L[f_L^0])_L$ is a minimizing sequence of $\T$. By Theorem \ref{free_vac} we infer $\nu_L[f_L^0]\to \bar f$ strongly in $L^2(B(0,\Lambda))$. Recall now that
$$f_L^0(p)=\alp\cdot A^0_L(p)+\beta b^0_L(p),\qquad \bar f(p)=\alp\cdot\omega_p\bar f_1(|p|)+\beta \bar f_0(|p|)$$
which easily implies that, more precisely, $\nu_L[A^0_L]\to \bar f_0(|\cdot|)$ and $\nu_L[b^0_L]\to \omega_\cdot\bar f_1(|\cdot|)$ strongly in $L^2(B(0,\Lambda))$.

Using the equation \eqref{eq_f_L} fulfilled by $f_L^0$, we are now able to prove a better convergence:
\begin{lemma}\label{lemma_estim_D} One has
\begin{equation}
\lim_{L\to\ii}\norm{\cD^0_L-\cD^0}_{\gS_\ii(\bT_L)}=\lim_{L\to\ii}\left(\sup_{p\in\Gamma_\Lambda^L}\left|\cD^0_L(p)-\cD^0(p)\right|\right)= 0,
\label{limit_sup_1}
\end{equation}
\begin{equation}
\lim_{L\to\ii}\norm{\gamma^0_L-\gamma^0}_{\gS_\ii(\bT_L)}=\lim_{L\to\ii}\left(\sup_{p\in\Gamma_\Lambda^L}\left|f^0_L(p)-\bar f(p)\right|\right)= 0,
\label{limit_sup_3}
\end{equation}
\begin{equation}
\lim_{L\to\ii}\norm{\cP^0_L-\cP^0}_{\gS_\ii(\bT_L)}=\lim_{L\to\ii}\left(\sup_{p\in\Gamma_\Lambda^L}\left|\cP^0_L(p)-\cP^0(p)\right|\right)= 0.
\label{limit_sup_2}
\end{equation}
Hence, for $L$ large enough,
\begin{equation}
\forall p\in\Gamma_\Lambda^L,\quad |\cD^0_L(p)|^2\geq |p|^2+\frac{m_0^2}{4}.
\label{estim_D_box}
\end{equation}
\end{lemma}
\begin{proof}
Since \eqref{limit_sup_1} easily implies \eqref{limit_sup_3}, \eqref{limit_sup_2} and \eqref{estim_D_box}, we only have to prove that
\begin{equation}
\sup_{p\in\Gamma_\Lambda^L}\left|\frac{(2\pi)^{5/2}\pi}{L^3}\sum_{q\in\Gamma_\Lambda^L}A^0_L(p-q)\widehat{W_L}(q)-\int_{B(0,\Lambda)}\frac{\omega_{p-q}\bar f_1(|p-q|)}{|q|^2}dq\right|\to0
\label{estim_1}
\end{equation}
and that
\begin{equation}
\sup_{p\in\Gamma_\Lambda^L}\left|\frac{(2\pi)^{5/2}\pi}{L^3}\sum_{q\in\Gamma_\Lambda^L}b^0_L(p-q)\widehat{W_L}(q)-\int_{B(0,\Lambda)}\frac{\bar f_0(|p-q|)}{|q|^2}dq\right|\to0
\label{estim_2}
\end{equation}
as $L\to\ii$. We only treat \eqref{estim_2}: \eqref{estim_1} is obtained by the same arguments. This is easily done by noting that for instance
\begin{multline*}
\left|\frac{(2\pi)^{5/2}\pi}{L^3}\sum_{\substack{q\in\Gamma_\Lambda^L\\ |q|\leq \eta}}b^0_L(p-q)\widehat{W_L}(q)-\int_{B(0,\eta)}\frac{\bar f_0(|p-q|)}{|q|^2}dq\right|\\
\leq \frac{1}{2}\left(\frac{2\pi^2\mu}{L}+\left(\frac{2\pi}{L}\right)^3\sum_{\substack{q\in\Gamma_\Lambda^L\\ q\neq 0,\ |q|\leq \eta}}\frac{1}{|q|^2}+\int_{B(0,\eta)}\frac{dq}{|q|^2}\right)
\end{multline*}
due to $|b^0_L(p)|\leq 1/2$ and $|\bar f_0(p)|\leq 1/2$, and an easy estimate for the rest of the form
\begin{align*}
& \left|\frac{(2\pi)^{5/2}\pi}{L^3}\sum_{\substack{q\in\Gamma_\Lambda^L\\ |q|> \eta}}b^0_L(p-q)\widehat{W_L}(q)-\int_{B(0,\Lambda)\setminus B(0,\eta)}\frac{\bar f_0(|p-q|)}{|q|^2}dq\right|\\
& \quad \leq\left|\int_{B(0,\Lambda)\setminus B(0,\eta)}\left(\nu_L[b^0_L](p-q)\nu_L[1/|\cdot|^2](q) - \frac{\bar f_0(|p-q|)}{|q|^2}\right)dq \right|+O(1/L)\\
& \quad\leq C\norm{\nu_L[b^0_L]-\bar f_0}_{L^2(B(0,\Lambda))}+C'\norm{\nu_L[1/|\cdot|^2]-1/|\cdot|^2}_{L^2(B(0,\Lambda)\setminus B(0,\eta))}+O(1/L)
\end{align*}
which converges to 0 as $L\to\ii$, due to the strong convergence of $\nu_L[b^0_L]$ to $\bar f_0$ in $L^2(B(0,\Lambda))$.
\end{proof}

\medskip

\noindent {\bf $\bullet$ Step 6: }{\it $\gamma_L^0=f_L^0(-i\nabla)$ is the unique global minimizer of $\E_L^0$ on $\cG_\Lambda^L$ for $L$ large enough}.
 Using \eqref{estim_D_box}, we now show that $\gamma_L^0=\cP_L^0-I_\Lambda^L/2$ is the unique global minimizer of $\E^0_L$. Indeed, using again ideas from \cite{BBHS,HLS1}, we write for any $\gamma\in\cG_\Lambda^L$,
\begin{eqnarray*}
\E_L^0(\gamma) & = & \E_L^0(\gamma_L^0)+ \tr(\cD^0_LQ)-\frac{\alpha}{2}\iint_{(\bT_L)^2}|Q(x,y)|^2W_L(x-y)dx\,dy\\
 & &\qquad + \frac{\alpha}{2}D_L(\rho_Q,\rho_Q)
\end{eqnarray*}
where we have used that, by construction,
$\rho_{\gamma_L^0}(x)=0$
for any $x\in\R^3$, and where $Q=\gamma-\gamma_L^0$. Notice that $Q$ satisfies the inequality
$$-\cP_L^0\leq Q\leq 1-\cP_L^0$$
which now implies as in \cite{BBHS,HLS1}
$$0\leq \tr(|\cD^0_L|Q^2)\leq \tr(\cD^0_LQ).$$
By Lemma \ref{prop_Coulomb}, we can estimate
\begin{eqnarray*}
\frac{\alpha}{2}\iint_{(\bT_L)^2}|Q(x,y)|^2W_L(x-y)dx\,dy & \leq & \frac{\alpha}{2}C_\Lambda^L(m_0/2)\tr(\sqrt{-\Delta+m_0^2/4}Q^2)\\
& \leq & \frac{\alpha}{2}C_\Lambda^L(m_0/2)\tr(|\cD^0_L|Q^2)
\end{eqnarray*}
due to \eqref{estim_D_box}, and therefore
$$\tr(\cD^0_LQ)-\frac{\alpha}{2}\iint_{(\bT_L)^2}|Q(x,y)|^2W_L(x-y)dx\,dy\geq \left(1-\frac{\alpha C_\Lambda^L(m_0/2)}{2}\right)\tr(|\cD^0_L|Q^2)$$
On the other hand, $D_L(\rho_Q,\rho_Q)\geq0$ and therefore
\begin{equation}
\E_L^0(\gamma)  \geq \E^0_L(\gamma_L^0)+\left(1-\frac{\alpha C_\Lambda^L(m_0/2)}{2}\right)\tr(|\cD^0_L|(\gamma_L^0-\gamma)^2)\label{estim_b_ener_L}
\end{equation}
For $L$ large enough one has
$$1-\frac{\alpha C_\Lambda^L(m_0/2)}{2}>0,$$
since $0\leq\alpha<4/\pi$ and $\lim_{L\to\ii}C_\Lambda^L(m_0/2)=C_\Lambda(m_0/2)\leq\pi/2$ by Lemma \ref{prop_Coulomb}. Therefore, by Lemma \ref{lemma_estim_D} and \eqref{estim_b_ener_L},
$$\E_L^0(\gamma)\geq \E^0_L(\gamma_L^0)+\frac{m_0\left(1-{\alpha C_\Lambda^L(m_0/2)}/{2}\right)}{2}\norm{\gamma_L^0-\gamma}_{\gS_2(\gH_\Lambda^L)}^2$$
which implies that $\gamma_L^0$ is the unique minimizer of $\E_L^0$ on $\cG_\Lambda^L$ and ends the proof of Theorem \ref{thermo1}. \qed

\subsection{Proof of Theorem \ref{thermo1b}}\label{proof_thermo1b}
For the sake of simplicity, we only treat the case where
$$\tilde{\mathbb H}_L^2=\sum_{i\geq1} (D^0)_{x_i}+\sum_{1\leq i<j}W_L(x_i-x_j)$$
is chosen instead of $\mathbb{H}^0_L$, the arguments being easy to extend to $\tilde{\mathbb H}_L^1$.

The energy of a Hartree-Fock state of density matrix $\Theta$ (a self-adjoint operator acting on $\gH_\Lambda^L$ such that $0\leq\Theta\leq I_\Lambda^L$) is now simply $\E_L^0(\Theta)$ instead of \eqref{QED_energy} (recall that $\E^0_L$ is defined in \eqref{QED_DF}). Therefore, restricting this energy to translation-invariant operators, the minimization problem to be solved for a fixed $L$ is
\begin{equation}
\tilde E_L:=\min\left\{\E_L^0(\Theta)\ \big|\ \Theta^*=\Theta,\  0\leq\Theta\leq I_\Lambda^L,\ \exists \xi\ |\  \Theta=\xi(-i\nabla)\right\}.
\end{equation}
Since $\gH_\Lambda^L$ is finite dimensional, there exists a minimizer $\Theta_L(x,y)=(2\pi)^{-3/2}\xi_L(x-y)$ of this energy, where
$$\xi_L(x)=(2\pi/L)^{3/2}\sum_{k\in\Gamma_\Lambda^L}{\Theta}_L(k)e_k(x).$$
Computing the energy of $\Theta_L$, we obtain
\begin{equation}
\E_L^0(\Theta_L) = \sum_{k\in\Gamma_\Lambda^L}\tr_{\C^4}(D^0(k)\Theta_L(k))-L^3\frac{\alpha}{2(2\pi)^3}\int_{\bT_L}|\xi_L(x)|^2W_L(x)dx
 +\frac{\alpha}{2}\mu L^5(\rho_{\Theta_L})^2.
\label{energy_other}
\end{equation}
Notice that
$$\left|\sum_{k\in\Gamma_\Lambda^L}\tr_{\C^4}(D^0(k)\Theta_L(k))\right|\leq 2\sqrt{\Lambda^2+m_0^2}\;|\Gamma_\Lambda^L|$$
since $0\leq \Theta_L(k)\leq I_{\C^4}$ for any $k$, and that
\begin{eqnarray*}
 \int_{\bT_L}|\xi_L(x)|^2W_L(x)dx & \leq & C_\Lambda^L(m_0)\pscal{|D^0|\xi_L,\xi_L}\\
 & = & C_\Lambda^L(m_0)\frac{(2\pi)^{3}}{L^3}\sum_{k\in\Gamma_\Lambda^L}|D^0(k)|\tr_{\C^4}(\Theta_L(k)^2)\\
 & \leq  & 4(2\pi)^{3}C_\Lambda^L(m_0)\sqrt{\Lambda^2+m_0^2}\frac{|\Gamma_\Lambda^L|}{L^3}
\end{eqnarray*}
by Lemma \ref{prop_Coulomb}. This implies
$$L^{-3}\left|\sum_{k\in\Gamma_\Lambda^L}\tr_{\C^4}(D^0(k)\Theta_L(k))-L^3\frac{\alpha}{2(2\pi)^3}\int_{\bT_L}|\xi_L(x)|^2W_L(x)dx\right|\leq M$$
for some uniform constant $M$. Hence, using $\E_L^0(\Theta_L)\leq0$,
we obtain from \eqref{energy_other}
$$\frac{\alpha}{2}\mu L^2(\rho_{\Theta_L})^2\leq M$$
and $\lim_{L\to\ii}\rho_{\Theta_L}=0$. To end the proof of Theorem \ref{thermo1b}, we notice that
\begin{eqnarray*}
\norm{\xi_L}_{L^\ii(\cC_L)}=\norm{\xi_L}_{L^\ii(\R^3)} & \leq &  \frac{(2\pi)^{3/2}}{L^3}\sum_{k\in\Gamma_\Lambda^L}|\Theta_L(k)|\\
 & \leq & \frac{C(2\pi)^{3/2}}{L^3}\sum_{k\in\Gamma_\Lambda^L}\tr_{\C^4}\Theta_L(k)\\
& \leq & C(2\pi)^{3/2}\rho_{\Theta_L}
\end{eqnarray*}
where we have used that $\Theta_L\geq 0$ and that $|M|=\tr_{\C^4}(M^*M)^{1/2}\leq C\tr_{\C^4}(|M|)$ for some constant $C$. This proves \eqref{limit_pL}.\qed

\begin{remark} Passing to the weak limit in the energy, it is also easy to prove that $\lim_{L\to\ii}{\tilde E_L}/{L^3}=0$.
\end{remark}

\subsection{Proof of Theorem \ref{thermo2}}\label{proof_thermo2}
\noindent {\bf $\bullet$ Step 1: }{\it Construction of a minimizing sequence of finite rank operators for $\E_{\rm BDF}^\phi$}. In order to prove an upper bound for the energy difference $E_L(\phi)-E_L(0)$, we need to use a trial state. To simplify matter, we shall use a finite rank operator.

\begin{lemma}
\label{approx_finite_rank}
Assume that $0\leq\alpha<4/\pi$, $\Lambda>0$, $m_0>0$ and that $n\in\cC$. Then there exists a sequence $(Q_k)\subset\cQ_\Lambda$ of finite rank operators such that $Q_k+\cP^0_-$ is a projector for any $k$, and
$$\lim_{k\to\ii}\E_{\rm BDF}^\phi(Q_k)=\min_{\cQ_\Lambda}\E_{\rm BDF}^\phi.$$
\end{lemma}
\begin{proof}
We know from Theorem \ref{BDF_min} that $\E_{\rm BDF}^\phi$ possesses a minimizer $\bar Q=\bar\cP_--\cP^0_- \in\cQ_\Lambda$ such that $\bar\cP_-$ is a solution of the self-consistent equation
$$\bar\cP_-=\chi_{(-\ii;0)}\left(\cD^0+\phi_{\bar Q}-R_{\bar Q}\right)$$
where
$$\phi_{\bar Q}:=\alpha(\rho_{\bar Q}-n)\ast\frac{1}{|\cdot|},\qquad R_{\bar Q}:=\alpha\frac{\bar Q(x,y)}{|x-y|}.$$
We know from \cite{HLS1,HLS2} that $\nabla\phi_{\bar Q}\in L^2(\R^3)$, $\phi_{\bar Q}\in L^6(\R^3)$ and $R_{\bar Q}\in \gS_2(\gH_\Lambda)$.
Now, let be $(\phi_k)$ and $(R_k)$ two sequences such that
\begin{enumerate}
\item $\phi_k\in L^1(\R^3)\cap H^1(\R^3)$ for any $k$, $\nabla\phi_k\to \nabla \phi_{\bar Q}$ in $L^2$ and $\phi_k\to\phi_{\bar Q}$ in $L^6(\R^3)$ as $k\to\ii$;
\item $R_k\in\gS_1(\gH_\Lambda)$ for any $k$ and $R_k\to R_{\bar Q}$ in $\gS_2(\gH_\Lambda)$ as $k\to\ii$.
\end{enumerate}
We define $P'_k:=\chi_{(-\ii;0)}\left(\cD^0+\phi_k-R_k\right)$ and $Q'_k:=P'_k-\cP^0_-$. For the sake of simplicity, we assume that $0$ is not in the spectrum of $\cD^0+\phi_{\bar Q}-R_{\bar Q}$, the following arguments being easily adapted to the other case. Then $0\notin\sigma(\cD^0+\phi_k-R_k)$ for $k$ large enough and we can use Cauchy's formula like in \cite{HS,HLS1} to obtain
\begin{eqnarray}
Q'_k & = & -\frac1{2\pi}\int_{-\ii}^\ii\left(\frac{1}{\cD^0+\phi_k-R_k+i\eta}-\frac{1}{\cD^0+i\eta}\right)d\eta\nonumber\\
 & = & \frac1{2\pi}\int_{-\ii}^\ii\frac{1}{\cD^0+\phi_k-R_k+i\eta}(\phi_k-R_k)\frac{1}{\cD^0+i\eta}d\eta \label{Cauchy}
\end{eqnarray}
which implies $Q'_k\in\gS_1(\gH_\Lambda)$ for
\begin{multline*}
\norm{\frac{1}{\cD^0+\phi_k-R_k+i\eta}(\phi_k-R_k)\frac{1}{\cD^0+i\eta}}_{\gS_1(\gH_\Lambda)}\\  \leq \frac{1}{\sqrt{\epsilon^2+\eta^2}\sqrt{m_0^2+\eta^2}}\norm{\phi_k-R_k}_{\gS_1(\gH_\Lambda)}
\end{multline*}
since $|\cD^0+\phi_k-R_k|\geq\epsilon>0$ and thanks to \eqref{propD3}. Notice that $\phi_k\in\gS_1(\gH_\Lambda)$ since, denoting by $\1_\Lambda$ the characteristic function of the ball $B(0,\Lambda)$,
\begin{eqnarray*}
\norm{\phi_k}_{\gS_1(\gH_\Lambda)} & = & \norm{\1_\Lambda(-i\nabla)\phi_k\1_\Lambda(-i\nabla)}_{\gS_1(\gH_\Lambda)}\\
 & \leq & \norm{\1_\Lambda(-i\nabla)|\phi_k|^{1/2}}_{\gS_2(\gH_\Lambda)}^2\\
 & \leq & C\norm{\1_\Lambda}_{L^2}^2\norm{\phi_k}_{L^1}
\end{eqnarray*}
by the Kato-Seiler-Simon inequality, see \cite{SeSi} and \cite[Theorem 4.1]{Simon}, or the appendix of \cite{H}.

We now show that $\lim_{k\to\ii}\E_{\rm BDF}^\phi(Q'_k)=\E_{\rm BDF}^\phi(\bar Q)$. Due to the results of Klaus-Scharf \cite{KS} or the estimates of \cite{HLS1}, we already know that $Q'_k\to\bar Q$ in $\gS_2(\gH_\Lambda)$.  Since $\E_{\rm BDF}^\phi$ is strongly continuous \cite{HLS1} for the norm
\begin{equation}
\norm{Q}:=\left(\norm{Q}_{\gS_2(\gH_\Lambda)}^2+\norm{\rho_Q}_\cC^2\right)^{1/2},
\label{norm_Q_HLS}
\end{equation}
it therefore remains to prove that $\rho_{Q'_k}\to \rho_{\bar Q}$ in $\cC$. Notice that due to the ultraviolet cut-off, $\rho_{Q'_k}\to\rho_{\bar Q}$ strongly in $L^2$ (see, e.g., \cite[Formula $(9)$]{HLS1}). Expanding \eqref{Cauchy}, we obtain
\begin{multline}
Q'_k=-\frac{1}{2\pi}\sum_{J=1}^5\int_{-\ii}^\ii d\eta \left(\prod_{j=1}^J\frac{1}{\cD^0+i\eta}(R_k-\phi_k)\right)\frac{1}{\cD^0+i\eta}\\
 -\frac{1}{2\pi}\int_{-\ii}^\ii d\eta \left(\prod_{j=1}^6\frac{1}{\cD^0+i\eta}(R_k-\phi_k)\right)\frac{1}{\cD^0+\phi_k-R_k+i\eta}
\label{exp_Cauchy}
\end{multline}
As proved in \cite[Section 4.3.3]{HLS1}, the first sum of the r.h.s. of \eqref{exp_Cauchy} is continuous for the norm \eqref{norm_Q_HLS} when $\nabla\phi_k\to\nabla\phi_{\bar Q}$ in $L^2(\R^3)$, $\phi_k\to\phi_{\bar Q}$ in $L^6(\R^3)$ and $R_k\to R_{\bar Q}$ in $\gS_2(\gH_\Lambda)$ (notice that the proof of \cite{HLS1} has to be adapted to the case where $D^0$ is replaced by $\cD^0$, which is an easy task). It therefore remains to show that the density associated with the last term of \eqref{exp_Cauchy} is also continuous for the $\cC$ norm as $k\to\ii$. The only non trivial term is
$$\int_{-\ii}^\ii d\eta \left(\prod_{j=1}^6\frac{1}{\cD^0+i\eta}\phi_k\right)\frac{1}{\cD^0+\phi_k-R_k+i\eta}$$
which indeed is continuous for the $\gS_1(\gH_\Lambda)$ topology by estimates of the type
\begin{align*}
& \norm{\left(\prod_{j=1}^6\frac{1}{\cD^0+i\eta}\phi_k\right)\frac{1}{\cD^0+\phi_k-R_k+i\eta}}_{\gS_1(\gH_\Lambda)}\\
& \qquad\leq C \norm{\frac{1}{\cD^0(p)+i\eta}}_{L^6}^6\norm{\phi_k}_{L^6}^6\frac{1}{\sqrt{\epsilon^2+\eta^2}}\\
& \qquad\leq C \norm{\frac{1}{D^0(p)+i\eta}}_{L^6}^6\norm{\phi_k}_{L^6}^6 \frac{1}{\sqrt{\epsilon^2+\eta^2}}\\
& \qquad=C(m_0^2+\eta^2)^{-3/2}(\epsilon^2+\eta^2)^{-1/2}\norm{(1+p^2)^{-1/2}}_{L^6}^6\norm{\phi_k}_{L^6}^6.
\end{align*}
This means that the associated density $(\rho_k)$ converges in $L^1$. Therefore the Fourier transform $(\widehat{\rho_k})$ converges in $L^\ii$ and has a compact support by assumption. Hence $(\rho_k)$ converges in $\cC$.

As a conclusion, $Q'_k\to \bar Q$ as $k\to\ii$ for the norm $\norm{Q}:=(\norm{Q}_{\gS_2(\gH_\Lambda)}^2+\norm{\rho_Q}_\cC^2)^{1/2}$. It is proved in \cite[Section 5.1]{HLS2} that $\E_{\rm BDF}^\phi$ is strongly continuous for this norm, which implies now $\lim_{k\to\ii}\E_{\rm BDF}^\phi(Q'_k)=\E_{\rm BDF}^\phi(\bar Q)$. Hence, we have constructed a minimizing sequence of trace-class operators in $\cQ_\Lambda$. As a last step, we now approximate each $Q'_k$ by a sequence of finite rank operators (since $k$ is fixed, we shall denote for simplicity $Q'=P'-\cP^0_-$ instead of $Q'_k$).

This will be done by using the following decomposition\footnote{The reader should compare this decomposition with the classical formula which gives the BDF state associated with the projector $P_k$ in the Fock space built with $\cP^0_-$, see, e.g., \cite[Formula $(10.96)$]{Thaller} where $A=\sum_{i\geq1}\lambda_i|u_i\rangle\langle v_i|$.} proved in \cite{HLS3}
\begin{equation}
P'=\sum_{n=1}^N|f_n\rangle\langle f_n|+\sum_{i=1}^\ii\frac{|u_i+\lambda_iv_i\rangle\langle u_i+\lambda_iv_i|}{1+\lambda_i^2},
\label{reduc_P}
\end{equation}
\begin{equation}
1-P'=\sum_{m=1}^M|g_m\rangle\langle g_m|+\sum_{i=1}^\ii\frac{|v_i-\lambda_iu_i\rangle\langle v_i-\lambda_iu_i|}{1+\lambda_i^2}.
\label{reduc_PP}
\end{equation}
where $(f_i)_{i=1}^N\cup(v_i)_ {i\geq1}$ is an orthonormal basis of $\cP^0_+\gH_\Lambda$, $(g_i)_{i=1}^M\cup(u_i)_{i\geq1}$ is an orthonormal basis of $\cP^0_-\gH_\Lambda$, and  $\sum_{i\geq1}\lambda_i^2<\ii$. Using \eqref{reduc_P} and \eqref{reduc_PP}, one sees that
\begin{multline}
Q' = \sum_{n=1}^N|f_n\rangle\langle f_n|-\sum_{m=1}^M|g_m\rangle\langle g_m|  +\sum_{i\geq 1}\frac{\lambda_i^2}{1+\lambda_i^2}\big(|v_i\rangle\langle v_i|-|u_i\rangle\langle u_i|\big)\\
+\sum_{i\geq 1}\frac{\lambda_i}{1+\lambda_i^2}\big(|u_i\rangle\langle v_i|+|v_i\rangle\langle u_i|\big)
\label{expansion_Q}
\end{multline}
which easily implies that $\sigma(Q')\cap(0;1)=\bigcup_{i\geq 1}\left\{\pm \frac{\lambda_i}{\sqrt{1+\lambda_i^2}}\right\}$. Recall that $Q'\in\gS_1(\gH_\Lambda)$ which implies $(\lambda_i)_i\in l^1(\R)$. Let us now define
$$P_{K}:=\sum_{n=1}^N|f_n\rangle\langle f_n|+\sum_{i=1}^K\frac{|u_i+\lambda_iv_i\rangle\langle u_i+\lambda_iv_i|}{1+\lambda_i^2}+\sum_{i=K+1}^\ii|u_i\rangle\langle u_i|.$$
The operator
\begin{multline}
Q_K=P_K-\cP^0_- = \sum_{n=1}^N|f_n\rangle\langle f_n|-\sum_{m=1}^M|g_m\rangle\langle g_m|  +\sum_{i= 1}^K\frac{\lambda_i^2}{1+\lambda_i^2}\big(|v_i\rangle\langle v_i|-|u_i\rangle\langle u_i|\big)\\
+\sum_{i=1}^K\frac{\lambda_i}{1+\lambda_i^2}\big(|u_i\rangle\langle v_i|+|v_i\rangle\langle u_i|\big)
\label{expansion_Q1}
\end{multline}
is a finite rank operator in $\cQ_\Lambda$ such that $Q_{K}\to Q'$ for the $\gS_1(\gH_\Lambda)$ topology as $K\to\ii$. Therefore $\lim_{K\to\ii}\E_{\rm BDF}^\phi(Q_{K})=\E_{\rm BDF}^\phi(Q')$.
This ends the proof of Lemma \ref{approx_finite_rank}.
\end{proof}

\medskip

\noindent {\bf $\bullet$ Step 2: }{\it Upper bound on the energy difference $E_L(\phi)-E_L(0)$}. Let be $\epsilon>0$ and $Q=P-\cP^0_-\in\cQ_\Lambda$ a finite rank operator of the form \eqref{expansion_Q1} for some $K$, such that
$$\E_{\rm BDF}^\phi(Q)\leq \min_{\cQ_\Lambda}\E_{\rm BDF}^\phi+\epsilon.$$
We now define, for $i=1,...,K$,
$$u_i^L(x)=\frac{(2\pi)^{3/2}}{L^3}\sum_{k\in\Gamma_\Lambda^L}\widehat{u_i}(k)e^{ik\cdot x}$$
and $(f_n^L)_{n=1}^N$, $(g_m^L)_{m=1}^M$, $(v_i^L)_{i=1}^K$ by similar formulas. We have, using the projector $\cP^0_L$ defined in Theorem \ref{thermo1},
$$\pscal{\cP^0_Lu_i^L,\cP^0_Lu_j^L}_{L^2(\bT_L)}=\left(\frac{2\pi}{L}\right)^3\sum_{k\in\Gamma_\Lambda^L}\pscal{\cP^0_L(k)\widehat{u_i}(k),\widehat{u_j}(k)}_{\C^4}$$
and so
\begin{equation}
\lim_{L\to\ii}\pscal{\cP^0_Lu_i^L,\cP^0_Lu_j^L}_{L^2(\bT_L)}=\int_{B(0,\Lambda)}\pscal{\cP^0_-(k)\widehat{u_i}(k),\widehat{u_j}(k)}_{\C^4}dk=\delta_{ij}.
\label{limit1}
\end{equation}
On the other hand, we also have
\begin{equation}
\lim_{L\to\ii}\norm{u_i^L-\cP^0_Lu_i^L}_{L^2(\bT_L)}^2=\pscal{\cP^0_+u_i,u_i}_{\gH_\Lambda}=0.
\label{limit2}
\end{equation}
Using the same type of behavior for $(f_n^L)_{n=1}^N$, $(g_m^L)_{m=1}^M$ and $(v_i^L)_{i=1}^K$ and the Gram-Schmidt orthonormalization procedure, we may therefore find an orthonormal system $(\tilde f_n^L)_{n=1}^N\cup(\tilde v_i^L)_{i=1}^K$ of $\cP^0_L\gH_\Lambda^L$ and an orthonormal system $(\tilde g_m^L)_{m=1}^M\cup(\tilde u_i^L)_{i=1}^K$ of $(1-\cP^0_L)\gH_\Lambda^L$ such that
\begin{equation}
\lim_{L\to\ii}\norm{\tilde u_i^L-u_i^L}_{L^2(\bT_L)}=0,\qquad \lim_{L\to\ii}\norm{\tilde v_i^L-v_i^L}_{L^2(\bT_L)}=0
\label{limit3}
\end{equation}
for all $i=1,...,K$, and
\begin{equation}
\lim_{L\to\ii}\norm{\tilde f_n^L-f_n^L}_{L^2(\bT_L)}=0,\qquad \lim_{L\to\ii}\norm{\tilde g_m^L-g_m^L}_{L^2(\bT_L)}=0.
\label{limit4}
\end{equation}
for all $n=1,...,N$ and $m=1,...,M$.

We now define our trial state by
\begin{multline}
\tilde Q_L= \sum_{n=1}^N|\tilde f^L_n\rangle\langle \tilde f^L_n|-\sum_{m=1}^M|\tilde g^L_m\rangle\langle \tilde g^L_m|  +\sum_{i= 1}^K\frac{\lambda_i^2}{1+\lambda_i^2}\big(|\tilde v^L_i\rangle\langle \tilde v^L_i|-|\tilde u^L_i\rangle\langle \tilde u^L_i|\big)\\
+\sum_{i=1}^K\frac{\lambda_i}{1+\lambda_i^2}\big(|\tilde u^L_i\rangle\langle \tilde v^L_i|+|\tilde v^L_i\rangle\langle \tilde u^L_i|\big)
\label{expansion_Q2}
\end{multline}
and $\tilde\gamma_L:=\gamma^0_L+\tilde Q_L$, where we recall that $\gamma^0_L=\cP^0_L-I_\Lambda^L/2$ is the unique translation-invariant minimizer of $\E_L^0$ defined in Theorem \ref{thermo1}. Notice that by construction, $\tilde Q_L$ satisfies
$$-\cP^0_L\leq \tilde Q_L\leq 1-\cP^0_L$$
and therefore $\tilde\gamma_L\in \cG_\Lambda^L$.
Let us now compute
\begin{eqnarray}
\E_L^\phi(\tilde\gamma_L)-E_L(0) & = & \E_L^\phi(\tilde\gamma_L)-\E_L^0(\gamma_L^0)\nonumber\\
 & = & \E_L^\phi(\gamma_L^0+\tilde Q_L)-\E_L^\phi(\gamma_L^0)\nonumber\\
 & = & \tr(\cD^0_L\tilde Q_L)-\int_{\bT_L}\phi_L(x)\rho_{\tilde Q_L}(x)\,dx+\frac\alpha2 D_L(\rho_{\tilde Q_L},\rho_{\tilde Q_L})\nonumber\\
 & & \qquad-\frac\alpha2 \iint_{(\bT_L)^2}|\tilde Q_L(x,y)|^2W_L(x-y)dx\,dy,\label{diff_trial}
\end{eqnarray}
where we have used that $\rho_{\gamma^0_L}\equiv0$. Passing to the limit in \eqref{diff_trial} and using \eqref{limit3} and \eqref{limit4}, one easily obtains
$$\lim_{L\to\ii}(\E_L^\phi(\tilde\gamma_L)-E_L(0))=\E_{\rm BDF}^\phi(Q)\leq \min_{\cQ_\Lambda}\E_{\rm BDF}^\phi+\epsilon$$
and therefore
$$\limsup_{L\to\ii}(E_L(\phi)-E_L(0))\leq \min_{\cQ_\Lambda}\E_{\rm BDF}^\phi.$$

\medskip

\noindent {\bf $\bullet$ Step 3: }{\it Lower bound on the energy difference $E_L(\phi)-E_L(0)$}. Since $\gH_\Lambda^L$ is finite-dimensional, there exists for any $L$ a minimizer $\gamma'_L$ of $\E_L^\phi$ on $\cG_\Lambda^L$. Using classical arguments already used in the Hartree-Fock theory \cite{Lieb,Bach,BLS} (see also \cite[Lemma 2]{HLS2}) and the positivity of $W_L$ (except on a set of measure zero), one easily shows that
$$\tilde\gamma'_L+I_\Lambda^L/2=\cP_L+\lambda|\phi\rangle\langle\phi|$$
where $\cP_L$ is an orthogonal projector, $\lambda\in[0;1]$ and $\phi\in{\rm ker}(\cD_{\gamma'_L})$ with
$$\cD_{\gamma'_L}=D^0+\alpha(\rho_{\gamma'_L}-n_L)\ast W_L-\alpha \gamma'_L(x,y)W_L(x-y)$$
Since then $\E_L^\phi(\cP_L-I_\Lambda^L)=\E_L^\phi(\gamma'_L)$, $\gamma_L:=\cP_L-I_\Lambda^L/2$ is also a global minimizer of $\E_L^\phi$ on $\cG_\Lambda^L$.

Let us now define $Q_L:=\gamma_L-\gamma_L^0=\cP_L-\cP^0_L$. Like in \eqref{diff_trial}, we have
\begin{eqnarray}
E_L(\phi)-E_L(0) & = & \E_L^\phi(\gamma_L)-\E_L^\phi(\gamma_L^0)\nonumber\\
 & = & \tr(\cD^0_L Q_L)-\int_{\bT_L}\phi_L(x)\rho_{Q_L}(x)\,dx+\frac\alpha2 D_L(\rho_{Q_L},\rho_{Q_L})\nonumber\\
 & & \qquad-\frac\alpha2 \iint_{(\bT_L)^2}|Q_L(x,y)|^2W_L(x-y)dx\,dy.\label{BDF_torus}
\end{eqnarray}
Therefore, arguing like in \cite{BBHS,HLS1,HLS2} and using \eqref{estim_D_box} and \eqref{Kato}, we infer
\begin{eqnarray}
E_L(\phi)-E_L(0) & \geq &  (1-\alpha C_\Lambda^L(m_0/2)/2)\tr\left(|\cD^0_L|(Q_L^{++}-Q_L^{--})\right)\nonumber\\
 & & \quad+\frac \alpha2 D_L(\rho_{Q_L}-n_L,\rho_{Q_L}-n_L)-\frac\alpha2 D_L(n_L,n_L)\nonumber\\
 & \geq & (1-\alpha C_\Lambda^L(m_0/2)/2)\tr\left(|\cD^0_L|Q_L^2\right)\nonumber\\
 & &  \quad+\frac \alpha2 D_L(\rho_{Q_L}-n_L,\rho_{Q_L}-n_L)-\frac\alpha2 D_L(n_L,n_L)\label{estim_below_BDF_torus}
\end{eqnarray}
where $Q_L^{++}:=(1-\cP^0_L)Q_L(1-\cP^0_L)\geq0$ and $Q_L^{--}:=\cP^0_LQ_L\cP^0_L\leq0$. Since $\lim_{L\to\ii}C_\Lambda^L(m_0/2)=C_\Lambda(m_0/2)\leq \pi/2$ and $0\leq\alpha<4/\pi$ by assumption, then $1-\alpha C_\Lambda^L(m_0/2)/2>0$ for $L$ sufficiently large. Using $\tr\left(|\cD^0_L|Q_L^2\right)\geq m_0/2\tr\left(Q_L^2\right)$ due to \eqref{estim_D_box}, we therefore deduce from \eqref{estim_below_BDF_torus} that
\begin{enumerate}
\item $Q_L(x,y)\1_{\cC_L}(x)\1_{\cC_L}(y)$ is bounded in $L^2((\R^3)^2)$ and therefore $\rho_{Q_L}(x)\1_{\cC_L}(x)$ is bounded in $L^2(\R^3)$ ;
\item $\rho_{|\cD^0_L|^{1/2}Q_L^{++}|\cD^0_L|^{1/2}}(x)\1_{\cC_L}(x)$ and $\rho_{-|\cD^0_L|^{1/2}Q_L^{--}|\cD^0_L|^{1/2}}(x)\1_{\cC_L}(x)$ are non-negative functions, uniformly  bounded in $L^1(\R^3)$.
\end{enumerate}
Thanks to the ultraviolet cut-off, we may therefore assume that
\begin{enumerate}
\item $Q_L(x,y)\1_{\cC_L}(x)\1_{\cC_L}(y)\wto \bar Q(x,y)$ weakly in $L^2((\R^3)^2)$ and uniformly on compact subsets of $\R^6$ ;
\item $\rho_{Q_L}(x)\1_{\cC_L}(x)\wto \rho_{\bar Q}(x)$ weakly in $L^2(\R^3)$ and uniformly on compact subsets of $\R^3$ ;
\item $\rho_{|\cD^0_L|^{1/2}Q_L^{++}|\cD^0_L|^{1/2}}(x)\1_{\cC_L}(x)\to \rho^+(x)\in L^1(\R^3)$ and $\rho_{-|\cD^0_L|^{1/2}Q_L^{--}|\cD^0_L|^{1/2}}(x)$ $\1_{\cC_L}(x)\to \rho^-(x)\in L^1(\R^3)$ uniformly on compact subsets of $\R^3$.
\end{enumerate}

Let us first show that $\bar Q\in\cQ_\Lambda$. Due to \eqref{limit_sup_2}, one easily obtains, passing to the weak limit,
$$-\cP^0_-\leq \bar Q\leq \cP^0_+.$$
Notice also that $|\cD^0_L|^{1/2}Q_L^{--}|\cD^0_L|^{1/2}(x,y)=|\cD^0_L|^{1/2}\cP^0_LQ_L\cP^0_L|\cD^0_L|^{1/2}(x,y)$ converges uniformly on compact subsets of $\R^6$ to $|\cD^0|^{1/2}\cP^0_-\bar Q\cP^0_-|\cD^0|^{1/2}(x,y)$, by \eqref{limit_sup_1} and \eqref{limit_sup_2}. Therefore, one obtains
$$\rho^+=\rho_{|\cD^0|^{1/2}\bar Q^{++}|\cD^0|^{1/2}}\in L^1(\R^3),\quad \rho^-=\rho_{-|\cD^0|^{1/2}\bar Q^{--}|\cD^0|^{1/2}}\in L^1(\R^3)$$
where, this time, $\bar Q^{--}=\cP^0_-Q\cP^0_-$ and $\bar Q^{++}=\cP^0_+Q\cP^0_+$, which shows that $\bar Q$ is $\cP^0_-$-trace class, $\bar Q\in\gS_1^{\cP^0_-}(\gH_\Lambda)$. Finally, \eqref{estim_below_BDF_torus} shows that $D_L(\rho_{Q_L}-n_L,\rho_{Q_L}-n_L)$ is bounded. Passing to the limit, we obtain $\rho_{\bar Q}-n\in\cC$ which implies $\rho_{\bar Q}\in\cC$ and as a conclusion $\bar Q\in\cQ_\Lambda$.
Hence, it remains to show that
$$\liminf_{L\to\ii}\left(E_L(\phi)-E_L(0)\right)\geq \E_{\rm BDF}^\phi(\bar Q).$$
This will imply the desired bound
$$\liminf_{L\to\ii}\left(E_L(\phi)-E_L(0)\right)\geq \min_{\cQ_\Lambda}\E_{\rm BDF}^\phi$$
since $\bar Q\in\cQ_\Lambda$.

By Fatou's lemma, we have
$$\liminf_{L\to\ii}D_L(\rho_{Q_L}-n_L,\rho_{Q_L}-n_L)\geq D(\rho_{\bar Q}-n,\rho_{\bar Q}-n)$$
and since obviously $\lim_{L\to\ii}D_L(n_L,n_L)= D(n,n)$ due to the approximation of integrals by Riemann sums for the continuous function $\widehat{n}$, it only remains to prove that
\begin{multline}
\label{estim_lim_inf}
\liminf_{L\to\ii}\left(\tr\left(|\cD^0_L|(Q_L^{++}-Q_L^{--})\right) -\frac\alpha2\iint_{(\bT_L)^2}|Q_L(x,y)|^2W_L(x-y)dx\,dy\right)\\
\geq \tr\left(|\cD^0|(\bar Q^{++}-\bar Q^{--})\right) -\frac\alpha2\iint_{\R^6}\frac{|\bar Q(x,y)|^2}{|x-y|}dx\,dy.
\end{multline}
This is shown by following exactly the method the authors of \cite{HLS2} used for the proof of their Theorem 1 (step 1), which we briefly outline now.

The idea of \cite{HLS2} is to use space cut-off functions $\eta_R$ and $\xi_R$ defined by $\eta_R(x)=\eta(|x|/R)$ and $\xi_R(x)=\xi(|x|/R)$ where $\eta,\xi\in\cC^\ii([0;\ii);[0;1])$ are such that $\eta^2+\xi^2=1$, $\eta(t)=1$ if $t\in[0;1]$ and $\eta(t)=0$ if $t\geq2$. Let us denote by $\eta_R^L$ the periodized cut-off function which is defined for $L$ large enough by the same formula as $\eta_R$ on $\cC_L$, and by $\xi_R^L=\sqrt{1-(\eta_R^L)^2}$.

Then, in order to reproduce the first step of the proof of \cite[Theorem 1]{HLS2}, the following analogue to \cite[Lemma 1]{HLS2} is needed:
\begin{lemma}\label{commutator}
We have $$\lim_{R\to\ii}\limsup_{L\to\ii}\norm{\,[|\cD^0_L|,\xi_R^L]\,}_{\gS_\ii(\gH_\Lambda^L)}=0.$$
\end{lemma}
\begin{proof}
We have
\begin{equation}
\norm{\,[|\cD^0_L|,\xi_R^L]\,}_{\gS_\ii(\gH_\Lambda^L)}\leq 2\norm{\,|\cD^0_L|-|\cD^0|\,}_{\gS_\ii(\gH_\Lambda^L)}+\norm{\,[|\cD^0|,\xi_R^L]\,}_{\gS_\ii(\gH_\Lambda^L)}.
\label{estim_comm}
\end{equation}
The first term of the right hand side of \eqref{estim_comm} tends to 0 as $L\to\ii$, due to Lemma \ref{lemma_estim_D}. By the regularity of $\cD^0(p)$ proved in Theorem \ref{free_vac}, there exists a constant $C$ such that
$$\big|\, |\cD^0(p)|-|\cD^0(q)|\,\big|\leq C|p-q|$$
when $p,q\in\Gamma_\Lambda^L$. This enables us to argue similarly to the proof of \cite[Lemma 1]{HLS2} and obtain a bound of the form
$$\norm{\,[|\cD^0|,\xi_R^L]\,}_{\gS_\ii(\gH_\Lambda^L)}\leq \frac{C}{L^3}\sum_{p\in(2\pi\Z^3)/L}\left|p\,\widehat{\xi_R^L}(p)\right|$$
and therefore
$$\limsup_{L\to\ii}\norm{\,[|\cD^0|,\xi_R^L]\,}_{\gS_\ii(\gH_\Lambda^L)}\leq C'\int_{\R^3}|r\widehat{\xi_R}(r)|\,dr=O(1/R),$$
which ends the proof of Lemma \ref{commutator}.
\end{proof}

One has
\begin{multline*}
\tr(|\cD^0_L|(Q_L^{++}-Q_L^{--}))= \tr(\eta_R^L|\cD^0_L|(Q_L^{++}-Q_L^{--})\eta_R^L)\\
+\tr(|\cD^0_L|\xi_R^L(Q_L^{++}-Q_L^{--})\xi_R^L)+\tr([\xi_R^L,|\cD^0_L|](Q_L^{++}-Q_L^{--})\xi_R^L),$$
\end{multline*}
On the other hand,
\begin{multline*}
\iint_{(\cC_L)^2}|Q_L(x,y)|^2W_L(x-y)dx\,dy=\iint_{\cC_L^2}\frac{\eta_R^2(x)\eta_{3R}^2(y)|Q_L(x,y)|^2}{|x-y|}dx\,dy\\
+\iint_{\cC_L^2}(\xi_R^L)^2(x)|Q_L(x,y)|^2W_L(x-y)dx\,dy+O(1/L)+O(1/R)
\end{multline*}
where we have used \eqref{conv_Coulomb}. Now, multiplying $-\cP_L^0\leq Q_L\leq 1-\cP^0_L$ by $\xi_R^L$ on both sides and using Kato's inequality \eqref{Kato} for $L$ large enough, we easily obtain, following \cite{HLS2}, that
$$\tr(|\cD^0_L|\xi_R^L(Q_L^{++}-Q_L^{--})\xi_R^L)-\frac\alpha2\iint_{\cC_L^2}(\xi_R^L)^2(x)|Q_L(x,y)|^2W_L(x-y)dx\,dy \geq0.$$
Hence,
\begin{multline}
\tr\left(|\cD^0_L|(Q_L^{++}-Q_L^{--})\right) -\frac\alpha2\iint_{\cC_L^2}|Q_L(x,y)|^2W_L(x-y)dx\,dy\\
\geq \tr(\eta_R^L|\cD^0_L|(Q_L^{++}-Q_L^{--})\eta_R^L)-\frac\alpha2\iint_{\cC_L^2}\frac{\eta_R^2(x)\eta_{3R}^2(y)|Q_L(x,y)|^2}{|x-y|}dx\,dy\\
+\tr([\xi_R^L,|\cD^0_L|](Q_L^{++}-Q_L^{--})\xi_R^L)+O(1/L)+O(1/R).
\end{multline}
The estimate \eqref{estim_lim_inf} is then easily obtained by passing to the limit first as $L\to\ii$ and then as $R\to\ii$, and using Lemma \ref{commutator}.

As a conclusion, we have shown that
$$\lim_{L\to\ii}(E_L(\phi)-E_L(0))=\min_{\cQ_\Lambda}\E_{\rm BDF}^\phi.$$
Due to our estimates, we also conclude that the weak limit $\bar Q$ of $Q_L$ satisfies $\E_{\rm BDF}^\phi(\bar Q)=\min_{\cQ_\Lambda}\E_{\rm BDF}^\phi$. Therefore, $\bar Q$ is a minimizer of $\E_{\rm BDF}$. This ends the proof of Theorem \ref{thermo2}.\qed








\bibliographystyle{plain}





\end{document}